\documentclass[pra,twocolumn,floatix,amssymb,showpacs,amsmath,superscriptaddress]{revtex4-1}

\usepackage[T1]{fontenc}
\usepackage[svgnames]{xcolor}
\usepackage[colorlinks=true,           
            linkcolor=DarkRed,
            urlcolor=Magenta,
            citecolor=Blue,
            anchorcolor=red]{hyperref}
\usepackage{amsmath, dsfont, bm}
\usepackage{physics,braket}
\usepackage{graphicx}
\usepackage{mathptmx}
\usepackage[makeroom]{cancel}

\begin{document}

\title{Motional Dynamical Decoupling for Matter-Wave Interferometry}
\author{Julen S. Pedernales}
\affiliation{Institut f\"ur Theoretische Physik und IQST, Albert-Einstein-Allee 11, Universit\"at Ulm, D-89081 Ulm, Germany} 
\author{Gavin W. Morley}
\affiliation{Department of Physics, University of Warwick, Gibbet Hill Road, Coventry CV4 7AL, United Kingdom}
\author{Martin B. Plenio}
\affiliation{Institut f\"ur Theoretische Physik und IQST, Albert-Einstein-Allee 11, Universit\"at Ulm, D-89081 Ulm, Germany} 

\begin{abstract}
Matter-wave interferometry provides a remarkably sensitive tool for probing minute forces and, potentially, the foundations of quantum physics by making use of interference between spatially separated matter waves. Furthering this development requires ever-increasing stability of the interferometer, typically achieved by improving its physical isolation from the environment. Here we introduce as an alternative strategy the concept of dynamical decoupling applied to spatial degrees of freedom of massive objects. We show that the superposed matter waves can be driven along paths in space that render their superposition resilient to many important sources of noise. As a concrete implementation, we present the case of matter-wave interferometers in a magnetic field gradient based on either levitated or free-falling nanodiamonds hosting a color center. We present an in-depth analysis of potential sources of decoherence in such a setup and of the ability of our protocol to suppress them. These effects include gravitational forces, interactions of the net magnetic and dipole moments of the diamond with magnetic and electric fields, surface dangling bonds, rotational degrees of freedom, Casimir-Polder forces, and diamagnetic forces. Contrary to previous analyses, diamagnetic forces are not negligible in this type of interferometers and, if not acted upon lead to small separation distances that scale with the inverse of the magnetic field gradient. We show that our motional dynamical decoupling strategy renders the system immune to such limitations while continuing to protect its coherence from environmental influences, achieving a linear-in-time growth of the separation distance independent of the magnetic field gradient. Hence, motional dynamical decoupling may become an essential tool in driving the sensitivity of matter-wave interferometry to the next level.
\end{abstract}

\maketitle

The quantum superposition principle, which follows from the linearity of the Schr\"odinger equation, allows a quantum mechanical system to occupy simultaneously several of the physical states available to it. An observable fingerprint of this, which applies not only to light but to matter as well~\cite{Broglie1923}, is the interference that the system will exhibit between the different states that contribute to such a superposition. For massive systems on the atomic scale, like electrons, atoms, ions, or molecules this has been tested and corroborated in laboratories all around the world \cite{Germer1927,Stern1930,Chu1991,Wineland1996}. However, as one scales up these systems, and with it their mass, the generation and observation of coherent superpositions becomes increasingly challenging. With a growing number of constituents the system becomes more and more sensitive to its environment, which manifests as noise acting to suppress the coherence in the superposition. Confronted with the absence of empirical evidence of macroscopic superpositions, the question arises: is this a result of our technical inability to generate a controllable environment where these superpositions can live sufficiently long to be observed, or are macroscopic systems described by a theory more fundamental than quantum mechanics that forbids the generation of macroscopic superpositions and from which the Schr\"odinger equation emerges as an effective model at the microscopic scale? 

Spontaneous collapse models offer a description of Nature for the latter scenario. They  postulate a modification of the  Schr\"odinger equation that incorporates nonlinear, stochastic terms, which account for a collapse of the wave function at macroscopic scales, while they ensure that the microscopic description of Nature remains unaffected~\cite{Ulbricht2013}. These models contain free parameters that quantify the strength and scale of such collapse mechanisms, and which, naturally, need to be determined empirically. A direct way to do so is to attempt to generate a coherent superposition of an object as macroscopic as possible. Then, using interferometric techniques one can assess the existence  and lifetime of such a coherent superposition, and use this information as a witness of spontaneous collapse models. However, there is a catch, the decoherence resulting from the interaction of the system with its uncontrolled environment, and that due to a postulated collapse model would be, in practice, indistinguishable. Therefore, the effort to test collapse models via interferometry, is in essence an effort to rule out any other known source of decoherence. If the system still decoheres, this could be due to an interaction of the system with its environment of which the experimenter is unaware or, indeed, have a fundamental origin. Instead, if the superposition survives for longer and longer times, this will set increasingly stringent bounds on the free parameters of possible collapse models. 

Mesoscopic matter-wave interferometers~\cite{Hornberger2013, Hornberger2014} have been implemented that bring objects with an increasing number of atoms to coherent superposition, based on BECs~\cite{Cornell1998}, fullerenes~\cite{Zeilinger1999}, or more recently macromolecules containing up to 10,000 amu~\cite{Tuxen2013,Arndt2013}. In parallel, a top-down approach is also being pursued with a variety of optomechanical systems proposed as potential macroscopic matter-wave interferometers~\cite{Tombesi1997, Knight1999,Bouwmeester2003,Black2004,Raizen2010,Raizen2011,Bouwmeester2012,Zoller2010,Cirac2011,Plenio2014}. Regarding this second approach, we are interested, in particular, in setups where the interaction of a strong magnetic field gradient with a nitrogen vacancy (NV) center is used to split the spatial wave function of the host nanodiamond, which can either be levitated or in free fall. This represents a promising platform, on the one hand, due to the remarkable coherence properties of NV centers even at room temperature~\cite{Walsworth2013,Taminiau2018}---in part, due to a mature understanding of NMR techniques applied to NV centers for the purposes of sensing and metrology~\cite{Hollenberg2013,Wu2016}---and, on the other hand, due to the recent and rapid development that the field has witnessed, demonstrating an increasing degree of controllability of NV centers in nanodiamonds that are levitated either by optical means~\cite{Vamivakas2013, Vamivakas2015, Vamivakas2017, Li2016a} or with ion traps~\cite{Benson2014, Hetet2017, Hetet2018, Quidant2018}. Not only that, a sizable spin-motion coupling is expected from such setups, which will significantly broaden the possibilities for NV based technologies. However, proposals for the use of nanodiamonds as matter-wave interferometers~\cite{Bose2013,Duan2013, Plenio2014, Kim2016, Morley2018} have so far been studied under idealised settings failing to consider the impact of a variety of physical features on their dynamics that are intrinsic to nanodiamond material. Notably, this includes the presence of diamagnetic forces on the system as well as significant sources of decoherence, such as the effect of electric and magnetic dipole moments, intrinsic nuclear and electron spins of the bulk and of the dangling bonds at the surface, or the presence of Casimir-Polder forces between the magnets and the diamond. In particular, the inclusion of diamagnetic forces in the analysis, far from being a mere act of refinement of the protocol, turns out to modify fundamentally the behaviour of the system, invalidating previously proposed schemes. Notably, the inclusion of diamagnetic forces leads to achievable spatial separation distances that are {\em inversely} proportional to the magnetic field gradient and grow at best linear in time, while their omission in previous studies led to separation distances that are proportional to the magnetic field gradient and which grow quadratically in time. Hence a careful analysis of all the relevant physical effects affecting the matter-wave interferometer is essential for assessing its potential.

Furthermore, the incorporation of sources of imperfections and noise in the analysis immediately raises the question of how to mitigate their impact while maintaining sensitivity to the desired signal; here, the relative phase accumulated in the spatial superposition. This challenge is reminiscent of that of quantum sensing using the electron spin of NV centers, where a wide variety of dynamical decoupling schemes have been developed that filter out slow noise while retaining sensitivity to a signal at a particular frequency \cite{Ryan2010,Lange2010,Naydenov2012,Cai2013,Mueller2014,Casanova2015}. In matter-wave interferometers, the misalignment of the interferometer with respect to gravity, variations of electric and magnetic fields etc. lead to significant and hard to control variations in the relative phase of the spatial wave function but are often slow or effectively static. Hence, introducing the concept of dynamical decoupling for the spatial degree of freedom in matter-wave interferometry as we do here can provide significant sensitivity improvements. In the light of all this, the design of new protocols is essential for bringing diamond based matter-wave interferometry closer to reality.

In this article, we present a novel experimental protocol for a matter-wave interferometer based on nanodiamonds. To this end, we introduce a pulse sequence that effects motional dynamics, and which, akin to pulsed dynamical decoupling, suppresses those system-environment interactions that are slow and linear: slow compared to the pulse spacing and linear in regard to the dependence of their potential energy in the position along of the superposition dimension. Furthermore, we show that our pulse sequence can be extended to suppress the effect of higher order interactions when these are weak compared to the energy of the system. Hence, the concept of motional dynamical decoupling and the specific protocols presented here reduce significantly the experimental requirements for successful matter-wave interferometry. As an added benefit, our proposed protocol makes use of resonant enhancement to address the impact of diamagnetic forces, achieving large separation between the components of the matter wave, with distances growing linearly in time, independent of the strength of the magnetic field gradient. Moreover, and in order to assess the practical feasibility of this novel protocol, we investigate in detail potential sources of dephasing that could act against the coherence of the superposition. We provide either mitigating strategies for some of these sources of noise or, failing that, a direction for material and experimental design efforts to overcome remaining challenges. 

The text is divided into two main sections. In the first section, we give a first-principles description of the system relevant to nanodiamond based matter-wave interferometry, and we introduce our experimental protocol, which realises dynamical decoupling of matter waves and takes into account the previously overlooked impact of diamagnetic forces on the interferometric scheme. The second section is devoted to the analysis of potential decoherence sources that could eventually degrade the visibility of the interference. We estimate their impact for realistic experimental parameters, the mitigating effect of our motional dynamical decoupling protocol and indicate how far the technology should push to make the interference of nanodiamonds visible for large spatial superpositions.

\section{Setup and protocol}
\label{sectionI}

Diamond being a diamagnetic material---that is, having negative magnetic susceptibility---is repelled by magnetic fields. Thus, when placed in an inhomogeneous magnetic field, a force directed towards the minimum of the field will be exerted on it. As a consequence, in the presence of a linear magnetic field gradient, and if no other force is acting on it, a diamond will behave as a harmonic oscillator with its equilibrium position at the zero of the field. This is the principle behind the magnetic trapping of diamagnetic objects, including the levitation of living organisms~\cite{Braunbek1939, Geim1997}. Now, if the diamond hosts an NV center, namely a spin-1 system with the gyromagnetic ratio of the electron, an additional force will be exerted on it originating from the interaction between the electron spin of the NV center and the magnetic field gradient. A spin aligned with the magnetic field will feel a force in the direction of the increasing magnetic field, while an antialigned spin will feel the same force in the opposite direction. In both cases, the effect of this force on the diamond is to shift the equilibrium position of the potential induced by the diamagnetic force, and thus, if the NV center is placed in a superposition of aligned and antialigned states, a superposition of two mechanical oscillators with two different equilibrium positions is obtained. Hence, without further measures, the achievable separation between the constituents of the superposition will be limited to twice the distance of the equilibrium points for any time. This is a novel view on the problem that departs from previous proposals, where diamagnetic forces were ignored and a quadratic in time and hence unlimited growth of the separation distance was predicted \cite{Kim2016,Morley2018}. In the following we will design a protocol that reduces the impact of diamagnetic forces and build our experimental proposal around it.

We consider an irregularly shaped diamond 
which contains a single negatively charged NV$^{-}$---hereinafter, just NV---that is localised at some random distance from the center of mass. Let us assume that the diamond is initially trapped in all its three translational degrees of freedom and cooled down to an internal temperature on the order of one Kelvin.  In the $x$-direction, a magnetic field gradient is applied, with the magnetic field assumed to be aligned with the axis of the NV inside the diamond. Now, two cases can be considered, that in which the diamond stays trapped and that in which the diamond falls freely along the z-direction due to the effect of gravity. These two scenarios are analogous regarding the dynamics in the x-direction as long as the trapping in this dimension occurs only due to the presence of the magnetic field gradient---recently, the trapping and cooling of nanodiamonds with magnetic traps has been demonstrated~\cite{Durso2016,Twamley2019}. For the clarity of the presentation, we describe here the case in which trapping forces are released, and the diamond falls in the $z$-direction, while the transversal magnetic field gradient in the $x$-direction is kept constant along the fall of the diamond.
The system is described by the Hamiltonian
\begin{multline}
\label{fullH}
H =  \underbrace{\frac{1}{2M} (\hat P^2_x + \hat P^2_y + \hat P^2_z)}_{\text{kinetic}} + \underbrace{M g \hat z}_{\text{gravitational}}\\
- \underbrace{\frac{\chi_{\rm V} V}{2 \mu_0} \hat{\bm B}^2}_{\text{diamagnetic}} - \underbrace{{\hat{ \bm \mu} \hat{\bm B}}}_{\text{NV-field}} + \underbrace{\hbar D \hat S_z^2}_{\text{zero-field}}.
\end{multline}
Here, $M$ is the mass of the diamond, $g$ the gravitational acceleration in the lab, $\chi_{\rm V}=- 2.2\cdot 10^{-5}$ and $V$ are, respectively, the volume magnetic susceptibility and the volume of the diamond, $\mu_0$ the vacuum permeability, and  $\bm \mu$ and $D = (2\pi)\ 2.8$ GHz are the magnetic moment and the zero-field splitting of the NV, respectively. The magnetic moment of the NV is given by ${\bm \mu} = - \hbar \gamma_{\text e} {\bm S}$, with $\gamma_{\rm e}= (2\pi)\ 28\  \mbox{GHz/T}$ the electronic gyromagnetic ratio and  $\bm S$ a vector containing the dimensionless spin-1 matrices for all three spatial dimensions. Let us consider a magnetic field of the form ${\bm B}  = B' x\  {\bm e}_z^{\text{(nv)}}$, where $B'$ gives the magnitude of the magnetic field gradient, and the unit vector ${\bm e}_z^{\text{(nv)}}$ is aligned with the direction of the NV axis.  According to Hamiltonian~(\ref{fullH}) the motion in the three Cartesian coordinates is uncoupled and, therefore, can be treated independently. Motion in the $y$-direction is free and will lead to a spreading of the initial wave function, in the $z$-direction the motion is accelerated due to the effect of gravity, and in the $x$-direction the motion corresponds to that of a forced harmonic oscillator. For our aim we are particularly concerned with the motion in the $x$-direction, described by the Hamiltonian
\begin{align}
\label{HamX}
\nonumber H_x &= \frac{1}{2M} \hat P^2_x - \frac{\chi_{\text V} V B'^2 }{2 \mu_0}  \hat x^2 +  \hbar \gamma_{\text e} B' \hat S_z \hat x \\
&= \hbar \omega \hat{a}^\dag \hat{a} + \hbar \lambda (a + a^\dag) \hat{S}_z,
\end{align}
where we have introduced the ladder operators $a$ and $a^\dag$, and defined the oscillation frequency $\omega = \sqrt{\frac{-\chi_{\text V}}{\rho_{\text D} \mu_0}} B'$ and spin-motion coupling strength $\lambda = \gamma_{\text e}\sqrt{\frac{\hbar}{2 M \omega}} B'$. Here, $\rho_{\text D} = 3510\ {\rm Kg}/{\rm m}^3$ is the mass density of diamond.

The Hilbert space of the NV can be spanned in terms of the eigenstates of the spin projection operator along the NV axis, $ \{ \ket{m_s} \} = \{ \ket{+}, \ket{-}, \ket{0} \}$, and the Hamiltonian reformulated in terms of displaced ladder operators associated to each of the NV spin states
\begin{equation}
\begin{split}
H_x =& \quad\hbar \omega a_+^\dag a_+ \ketbra{+}{+}  + \hbar \omega a_-^\dag a_-\ketbra{-}{-}  \\
&+ \hbar \omega (a^\dag a + \lambda^2/\omega^2)\ketbra{0}{0} .
\end{split}
\end{equation}
Here, $a_\pm = a \pm \lambda/\omega$ and we have shifted the total energy by a constant $\hbar \lambda^2/\omega$, with no loss of generality. The system is, therefore, a set of three harmonic oscillators revolving, with the same frequency, around three different equilibrium positions. In particular, we have that spin state $\ket{0}$ is associated to equilibrium position $x_{\rm eq}^0 = 0 $, while the equilibrium positions associated to NV states $\ket{\pm}$ are
\begin{equation}
x_{\rm eq}^\pm = \pm 2 x_0 \lambda / \omega,
\end{equation}
with $x_0 = \sqrt{\hbar / (2 M \omega)}$. 

Our goal is to generate a superposition with the greatest possible separation distance between the two superposed states. Let us assume that the nanodiamond is initially in an unspecified motional state $\rho_{\rm m}$ with the NV in state $\ket{0}$ when the diamond is dropped along the z-direction. Immediately after, a microwave pulse is applied to take the NV into a superposition of $\ket{\pm}$ states. The system transitions into a superposition of two components each oscillating around spatially separated equilibrium positions. An operator can be defined in the Heisenberg picture whose expectation value corresponds to the separation distance between these two oscillators, namely
\begin{align}
\delta \hat x (t) &= U^\dag_+ \hat x U_+ - U^\dag_- \hat x U_- \\
&= \Delta x_{\text{eq}} (\cos{\omega t} - 1 ),
\end{align}
where $U_{\pm} = e^{-i \omega a^\dag_\pm a_\pm t}$, and  
\begin{equation}
\label{sepeq}
\Delta x_{\text{eq}} = 4 x_0 \lambda/\omega = \frac{ 2 \hbar \gamma_{\text e} \mu_0}{- \chi_{\text V}} \frac{1}{ V B' }
\end{equation}
is the separation distance between the equilibrium positions of the two oscillators. Remarkably, the maximum separation distance is independent of the initial motional state, i.e. initial displacement from $x=0$ and velocity, and is determined only by the distance between the equilibrium positions. This indicates that ground state cooling of the motional degrees of freedom is not required for such a scheme. 

From Eq.~(\ref{sepeq}) we see that the smaller the magnetic field gradient the larger the separation one can achieve. This is one of the central results of this work for it contradicts previous proposals where diamagnetic forces were ignored and the separation distance was predicted to increase with the magnetic field gradient \cite{Kim2016,Morley2018}. Notice that for proposals with levitated nanodiamonds where the trapping is achieved by means other than the magnetic field gradient, the reachable separation distance will still grow with the magnetic field gradient as long as the diamagnetic confinement is weaker than the employed trapping mechanism. For optically trapped particles, the diamagnetic confinement will exceed the optical one only at unrealistically strong magnetic field gradients, and therefore neglecting the diamagnetic forces is justified in these cases~\cite{Bose2013,Duan2013}, albeit, for the same value of the magnetic field gradient, the reachable separation distances will be considerably smaller than in the case of pure diamagnetic trapping.

In order to estimate achievable separation distances, one should take into account that the oscillation frequency $\omega$ depends linearly on the magnetic field gradient. For a proper interferometric protocol, one should allow at least one oscillation to occur within the coherence time of the NV and the duration of the drop, such that the superposed states of diamond can come back to their initial positions in the x-axis, and the NV used to observe their interference. The desired oscillation period $T$ can be fixed by setting the value of the magnetic field gradient according to $B' = \sqrt{\frac{\mu_0 \rho_{\text D} }{- \chi_{\text V}}} 2 \pi /T$; then for a given oscillation period the maximum separation distance will be given by
\begin{equation}
\label{sepdisnoseq}
D_{\rm max} = 2 \Delta x_{\rm eq} = \frac{2 \hbar \gamma_{\rm e}}{\pi V} \sqrt{\frac{\mu_0}{\chi_{\rm V} \rho_{\rm D}}} T.
\end{equation}
The system will periodically split and reunite, with the amplitude of these oscillations depending linearly on their period $T$. Thus, the maximal separation distance is reached for the first time at half a period from the start of the experiment, and does not grow in time contrary to predictions of previous proposals~\cite{Kim2016,Morley2018}.

\subsection{A resonant amplification mechanism}
A harmonic oscillator subjected to a periodic driving force that is resonant with its oscillation frequency will experience a linear-in-time growth of its amplitude. This intuition will be used in the following to devise a scheme that makes the separation distance between the two superposed oscillators grow linearly in time. Remarkably, as an additional benefit, this  amplification protocol will lead to a symmetrisation of the paths of the two parts of the superposition which in turn will cancel uncontrolled phases due to spatially inhomogeneous potentials of unknown magnitude. This will be discussed in detail in section~\ref{section2}.

\begin{figure}[htbp]
\includegraphics[width=\columnwidth]{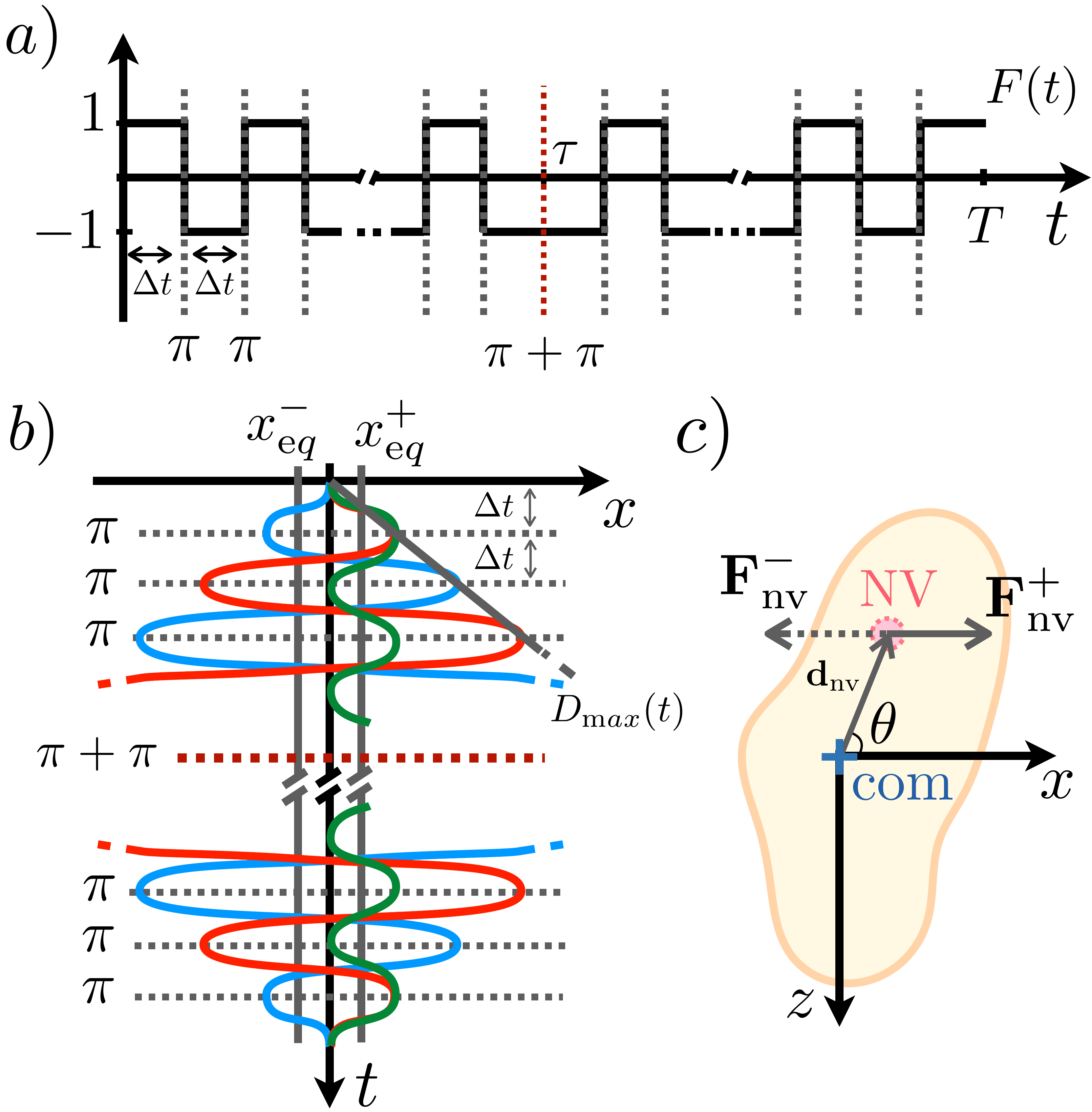}
\caption{{\bf Schematics of the setup and the dynamics.} (a) shows function $F(t)$ with vertical dashed lines indicating $\pi$-pulses. An additional $\pi$-pulse (red dashed line) is applied at time $\tau$, when the desired separation has been reached, in order to reverse the dynamics and bring the diamond back to its initial state at time $T = 2 \tau$. Therefore, at time $\tau$ two $\pi$-pulses are applied, or equivalently none. In (b) the amplified oscillations associated with spin up (down) are shown in red (blue). The amplitude of the oscillations grows linearly in time as $D_\text{max}(t)$, depicted in the figure with a diagonal gray line. Vertical lines indicate the equilibrium positions of the two oscillators. The oscillations corresponding to the spin up state in the absence of the pulse sequence are indicated in green. In (c) an irregularly shaped diamond is depicted in pale yellow, with an NV (pink) sitting at a position $\bm d_{\text{nv}}$ from the center of mass (blue). The forces acting on the NV due to the magnetic field gradient are also depicted. The rotation angle $\theta$ is measured as the angle between vectors $\bm{F}^+_{\text{nv}}$ and $\bm d_{\text{nv}}$.}
\label{fig:Scheme}
\end{figure}

Consider that we apply a series of $\pi$-pulses between states $ \ket{+}$ and $\ket{-}$ of the NV center while the diamond is falling. In the presence of these pulses, Hamiltonian~(\ref{HamX}) is modified as
\begin{equation}
\label{eq:Hamred}
H_{x + \text{MW}} = \hbar \omega a^\dag a + \hbar \lambda (a^\dag + a) S_z + \hbar \Omega(t) \sigma_x,
\end{equation}
with $\Omega(t)$ a time-dependent Rabi frequency, which has value 0 in between the pulses and some fixed value $\Omega = \pi/(2 \tau)$ while a pulse of duration $\tau$ is being applied, and $\sigma_x = \ketbra{+}{-} + \text{H.c.}$ We now move to an interaction picture with respect to the driving field $\hbar \Omega(t) \sigma_x$ and the free-energy part of the oscillator $\hbar \omega a^\dag a$. Furthermore, we assume that the pulses are instantaneous compared to any other frequencies in the system. This is a safe assumption as $\omega$ and the coupling strength $\lambda$ can be expected to be in the kHz regime, even for large magnetic field gradients of $10^5$ T/m. Our Hamiltonian is now found to be
\begin{equation}
\label{timedepHam}
H^\text{int}_{x + \text{MW}} =  \hbar F(t) \lambda S_z (a e^{-i \omega t}+ a^\dag e^{i \omega t}),
\end{equation}
where the effect of the $\pi$-pulses is to flip the sign of operator $S_z$. This is captured by the function $F(t)$, which is a piecewise constant function taking only values $1$ or $-1$, see Fig.~(\ref{fig:Scheme}\ a).

The unitary-evolution operator associated with the time-dependent Hamiltonian~(\ref{timedepHam}) can be exactly computed in the Magnus expansion, where terms of order $3$ and higher vanish identically.  Thus we have

\begin{equation}
U(t) = e^{\Omega^{(1)} + \Omega^{(2)}},
\end{equation} 
with
\begin{align}
\Omega^{(1)} &= (\alpha a - \alpha^* a^\dag) S_z\\
\Omega^{(2)} &= \text{const.} 
\end{align}
and 
\begin{equation}
\alpha = -i \lambda \int_0^t F(t') e^{-i \omega t'} dt'.
\end{equation}
The integral in $\alpha$ can be solved in the regions where F(t) is constant and the results added. This yields
\begin{eqnarray*}
\int_0^t F(t') e^{-i \omega t'} dt' =\sum_{n=0}^{N} \frac{(-e^{i \omega \Delta t})^n}{-i \omega} (e^{-i \omega \Delta t} - 1),\nonumber
\end{eqnarray*}
where N is the number of $\pi$-pulses and $\Delta t$ the interpulse spacing, such that the total time is $t = (N+1) \Delta t$. Clearly, the value of the integral is maximal when $\omega \Delta t = \pi$, that is to say when the pulse sequence is resonant with twice the frequency of the oscillator. In this situation, $\alpha$ takes the value
\begin{equation}
\alpha = - \frac{2 \lambda}{\pi} t.
\end{equation}
As a result the time evolution is given by the displacement operator 
\begin{equation}
\label{displacementOp}
U(t) = e^{(\alpha a - \alpha^* a^\dag)S_z},
\end{equation} 
with the displacement direction dependent on the state of the NV. Notice that $\alpha$ is real,  which indicates that the displacement will occur in the position quadrature, and that it depends linearly on time. The separation between the two superposed states will be given by $D_\text{max} = 2 x_o (\alpha + \alpha^*) = \frac{8}{\pi} x_0 \lambda t$. Remarkably, the quantity $x_0 \lambda$ does not depend on the magnetic field gradient, and the achievable separation distance is determined only by the volume of the particle and time,  
\begin{equation}
D_\text{max} /t = \frac{4 \hbar \gamma_{\text e}}{\pi V} \sqrt{\frac{\mu_0}{- \chi_{\text V} \rho_{\text D}}}.
\end{equation}
Once the desired separation has been reached, say at time $\tau$, an additional $\pi$-pulse can be applied followed by the time-reversed sequence of pulses to reverse the displacement and bring the diamonds together at time $T = 2 \tau$. In Fig.~(\ref{fig:Scheme} b) a scheme of the path is shown. 

By comparison with Eq.~({\ref{sepdisnoseq}}), we observe that the same separation distance is achieved in the absence of an amplification sequence, by fixing the magnetic field to the minimum value that allows one oscillation in time $T$. Thus, the aim of the amplification sequence is not to reach larger separation distances, but to allow us to achieve the same separation distances irrespective of the magnetic field gradient. This is not a marginal result. In general, working at higher magnetic fields will be preferable when considering realistic noise sources. This is because the same pulses that we use to generate the resonance can serve as a dynamical decoupling sequence to suppress the effect of non-Markovian noise sources. This may be used to extend  the coherence time of the NV and to suppress slow noise on the motional degrees of freedom (see sec. \ref{section2} for a detailed discussion), and in general a larger number of pulses is preferable in this respect, which in turn favours a large oscillation frequency and therefore a large magnetic field gradient. Notice, however, that the largest considerable magnetic field gradient is limited by the minimal distance between the diamond and the magnets that allows to neglect the effect of Casimir-Polder forces between the two, see Appendix~\ref{app:Casimir}.

Be it in the presence of a pulse sequence or not, smaller particles achieve larger separation distances, as in both cases the separation distances are inversely proportional to the volume of the oscillator. For a particle with the volume of a $1 \mu$m radius sphere, we find a separation that grows in time as $D_\text{max} /t~\approx~2.3~\times~10^{-8}~\frac{\text m}{\text s}$, while a particle with a radius of $230$~nm should result in a separation that grows as $D_\text{max}/t \approx 1.8~\times~10^{-6}$~m/s. For a free fall time of $T=0.5$~s, we set $\tau = 0.25$~s, which gives a maximal separation of $D_\text{max} \approx 5.75$~nm for a $1\  \mu$m radius particle and a separation of $D_\text{max} \approx 456$~nm if the radius of the particle is $230$~m. In the absence of a pulse sequence, this requires a magnetic field gradient of $B' = 181. 86$~T/m, while use of a pulse sequence allows us to obtain the same results also for larger values of the magnetic field gradient. 

Our scheme is not restricted to diamagnetically trapped particles and can be applied in general to any system that is described by a Hamiltonian of the form in Eq. (\ref{eq:Hamred}), that is, harmonically trapped particles with a spin degree of freedom that couples linearly to their motion. In particular, relevant to the topic of this article are nanodiamonds that are levitated by means other than diamagnetic forces, such as by optical forces or in Paul traps. In these cases, the confining forces will typically be stronger and therefore the reachable separation distances in the absence of a pulse sequence notably smaller, $D_{\rm max} = \frac{4 \hbar \gamma_e B'}{M \omega^2}$. Thus, such setups will strongly benefit from the design introduced here in order to go beyond this limitation and reach separation distances that grow linearly in time as $D_{\rm max} = \frac{4 \hbar \gamma_e B'}{M \omega \pi} t$. Finally,  it is noteworthy to mention that the discussed protocol can be analogously applied to transitions $\ket{0} \rightarrow \ket{\pm}$, albeit halving the reachable separation distance, as the equilibrium positions of the two superposed oscillators sit now at half the distance from each other.

\section{Sources of decoherence}
\label{section2}
The interferometric protocol works as follows. Initially, the diamond is trapped and cooled down to a low temperature thermal state, with its internal degree of freedom in state $\ket{0}$. A suitable combination of microwave pulses prepares the NV in the superposition state $\ket{+} + \ket{-}$ and subsequently the diamond is dropped across the transversal magnetic field gradient. The resonant pulse sequence described in the previous section results in a spin-dependent displacement of the center of mass of the diamond, which as a consequence splits in two spatially separated superposed states, with a separation distance that grows linearly in time. Once the desired separation has been reached, an additional $\pi$-pulse inverts the dynamics, reuniting the two superposed states of the diamond, ideally to a perfect overlap of their wave functions. At this moment, if the process was coherent the NV and the diamond will be found in a separable state, with the NV in an even superposition of states $\ket{\pm}$. A final
operation on the NV takes it back to state $\ket{0}$; then, the population of this state is measured. Any population leak into states $\ket{\pm}$ reveals a loss of coherence in the process. For example, if a collapse mechanism destroys the superposition at some point during the protocol, such that the state of the system collapses to, say, state $\ket{+}$($\ket{-}$) of the NV, the gate at the end of the protocol, which would send the NV into state $\ket{0}$ in ideal conditions, will instead populate state $\ket{-}$($\ket{+}$). Therefore, the observation of population in states other than $\ket{0}$ becomes the smoking gun for decoherence. However, other decoherence mechanisms that are not related to an underlying collapse model could have a similar effect. Therefore, it becomes crucial to discard these sources of noise, in order to be able to relate any observed loss of coherence to a fundamental origin.

In this section, we give an extensive list of system-environment interactions that could lead to a loss of coherence during the protocol. We divide these into two main categories. On the one hand,  (A) we consider those which introduce a relative phase between the two superposed states, such that, after the two paths are reunited, the spatial wave packets of the diamond still overlap and are to a good approximation in a product state with that of the NV, as in the ideal case, but a relative phase between states $\ket{\pm}$ of the NV has been picked up. Shot-to-shot variations of this phase will amount to a loss of the coherence information, with the expectation value of any operator being equivalent to that of having the NV in a mixture of states $\ket{\pm}$. The second type of incoherence that we consider (B) is that in which the two superposed states of the diamond do not perfectly overlap after they are brought together at the end of the protocol. The result will be that the NV and the motional degrees of freedom will not be in a separable state at the end of the protocol, which again will result in a population leakage into states $\ket{\pm}$.  

\subsection{Relative phases}

Relative phases between the two superposed states of the system arise because the potential energy integrated over the two paths of the nanodiamond are inequivalent. We will be working in a reference frame of the setup in which the x-direction is defined by the magnetic field gradient. In this frame, for an ideally working setup, the two superposed states will differ only in the $x$-coordinate. Consequently, we are primarily concerned with slowly varying potential energy terms that are inhomogeneous across this direction. Of special interest will be terms that depend linearly on $x$, ${V(x) = - F_0 x}$, because they are expected to be the leading order terms in any perturbation. 
We will consider the cases where these forces have a constant magnitude over time as well as the case in which their magnitude varies stochastically but slowly. It is a main result of this section that with a suitable pulse sequence their effect can be canceled. We will also consider the effect of potential energy terms that depend nonlinearly on $x$, and thus result in forces that are inhomogeneous along the superposition dimension. 

\subsubsection{Constant homogeneous forces}
In the presence of a potential energy term that is linear in $x$, the system Hamiltonian that rules the dynamics in the $x$ direction is transformed into 
\begin{equation}
H_x = \hbar \omega a^\dag a  + \hbar \lambda (a + a^\dag) S_z - F_0x_0 (a + a^\dag).
\end{equation}
The new, third, term can be absorbed by redefining the ladder operators as ${c^\# = a^\# - \frac{F_0x_0}{\hbar \omega}}$, which now account for the displacement of the equilibrium positions. In terms of these new ladder operators the Hamiltonian becomes
\begin{equation}
H_x = \hbar \omega c^\dag c + \hbar \lambda (c + c^\dag) S_z + \hbar \phi S_z,
\end{equation} 
with $\phi = \frac{2 \lambda x_0}{\hbar \omega}F_0$, making it evident that under a constant force of magnitude $F_0$, a relative phase $2\phi t$ will be picked up between the states $\ket{\pm}$ of the NV. Notice, that this is precisely the phase corresponding to the difference in potential energy between the equilibrium positions of the two oscillators, $2 \phi t = \Delta x_\text{eq} (F_0/\hbar) t$, see Eq.~(\ref{sepeq}).

For a nanodiamond of $230$ nm radius in a magnetic field gradient of $10^3$ T/m the equilibrium positions are separated by a distance $\Delta x_{\rm eq} \approx 43$ nm, such that for an experiment lasting a time on the order of $0.5$~s, forces as small as $10^{-24}$ N will already introduce a phase of $2 \phi t \approx 1$ rad. We emphasise once more that the separation distance of the superposition is independent of the magnetic field gradient; for the example above it would be approximately $470$ nm, which is sufficient to ensure that there is no overlap between the two components of the superposition. However, since the acquired relative phase is directly proportional to the separation distance between the equilibrium positions---not the separation distance between the paths---, and this is inversely proportional to the magnetic field gradient, a stronger magnetic field gradient should be preferable. 

In the following we consider the physical origin of forces of such a type. 

\paragraph{Misalignment with the gravitational field}

If the setup and the gravitational field are not well aligned, the gravitational force will acquire a component of magnitude $F_0 = M {\rm g} \sin \beta$ in direction $x$, $\beta$ being the angle between the gravitational force and the $z$-direction. A diamond of radius $230$~nm as the one discussed earlier has a mass $M \approx 10^{-16}$~Kg, which shows that a random misalignment of $10^{-9}$~rad would already result in a force on the order of $10^{-24}$~N, which as discussed above in an interference experiment lasting $0.5$~s would introduce a phase shift on the order of rads. Such a phase would destroy the interference unless held constant across the entire sequence of experiments which, without further measures, imposes daunting requirements on the stability of the orientation of the setup.


\paragraph{Magnetic moment of the diamond}

Apart from the NV, the diamond will contain additional spins which will contribute to giving it a net magnetic moment. Two dominant sources of these spins will be the nuclear spins of $^{13}$C atoms present in the lattice and the electronic spins originating either from the electrons in dangling bonds at the surface of the nanodiamond or from other defects in the lattice structure of the diamond, like for example P1 centers. A diamond of $230$~nm radius contains $\sim 10^{9}$ carbon atoms, which indicates that even for $^{13}$C concentrations as low as $10^{-6}$~\% the diamond will contain $\sim 10^3$ nuclear spins. On the other hand, the most optimistic estimates for the dangling bond density on the surface of diamond are on the order of $10^{-2}$ dangling bonds per nm$^2$ (estimated for single crystal bulk diamond grown under optimal conditions)~\cite{Oosterkamp2018}, which amounts to a total of $10^3$ electron spins on a $230$~nm diamond. All in all, one could expect that such a diamond will have a magnetic moment of $\mu_{\rm D} = \hbar 10^{3}  (\gamma_{\rm n} \expval{S^z}_{\rho^{\rm n}_{\rm th}}  + \gamma_{\rm e} \expval{S^z}_{\rho^{\rm e}_{\rm th}})$, where $S^z$ is the $z$ component of the spin-$1/2$ operator and the expectation value is taken over thermal states for the nuclei, $\rho^{\rm n}_{\rm th}$, and electrons, $\rho^{\rm e}_{\rm th}$. At a temperature of $1$~K and a magnetic field of $10^{-2}$~T this gives a magnetic moment of $\mu_{\rm D} \approx 10^{-22}\ \frac{\text{N m}}{\rm T}$. A magnetic field gradient of $10^2$~T/m acting on such a magnetic moment will exert a force on the order of $\sim 10^{-20}$~N, well above the the $10^{-24}$~N threshold established before. Shot-to-shot variations of the intrinsic magnetic moment of the diamond are to be expected due to thermal fluctuations of the spins that form it, which should scale with the square root of the number of spins, $\Delta \mu_{\rm D} = \hbar 10^{3/2}  (\gamma_{\rm n} \expval{S^z}_{\rho^{\rm n}_{\rm th}} + \gamma_{\rm e} \expval{S^z}_{\rho^{\rm e}_{\rm th}})$. Such variations would irremediably lead to a loss of coherence and consequently to a degradation of the visibility of the interference unless addressed by methods of dynamical nuclear polarisation transferring optically induced polarisation of the NV center to the surrounding spins \cite{Cai2013a,London2013,Schwartz2018} 

An alternative way to understand the origin of the relative phase introduced by the intrinsic magnetic moment of the diamond is in terms of the experienced Larmor frequencies along the two paths. The inhomogeneity of the magnetic field will clearly lead to the two superposed states picking up different phases.

\paragraph{Electric dipole moment of the diamond}

A charge neutral nanodiamond containing a negatively charged NV center will necessarily have an electric dipole moment, originated from the negative charge of the NV and the compensating positive charge, which will typically be located at a random position in the bulk of the diamond
\cite{Yao2018}. Continuing our analysis for a $230$ nm radius particle, let us assume that the compensating positive charge is located $200$ nm away from the NV center (note that for a uniform distribution of NV and charge we find as mean square distance $1.2 r^2$ in a spherical particle of radius $r$). This will cause an electric dipole moment of the order of $|\bm d | = 3.2\cdot 10^{-26}$ Cm and a force on the diamond of the order of $10^{-24}$ N in an electric field gradient of $30$ V/m$^2$ (which would for example be caused by a single electron at a distance of $500$ microns).

\subsubsection{A pulse sequence to suppress relative phases}
Here we show that relative phases due to constant homogeneous forces, such as those described above, can be suppressed by a suitable pulse sequence. In the presence of $\pi$-pulses the acquired phase due to constant homogeneous forces will now be given by 
\begin{equation}
2 \phi \int_0^T P(t') dt',
\end{equation}
where $P(t)$ is a function that flips between values $1$ and $-1$ after each $\pi$-pulse, and is in general different from the amplification sequence introduced in section~\ref{sectionI}, but can be made such that it serves both purposes. Notably, this integral can be made identically 0 if an odd number of $\pi$-pulses is applied with a fixed interpulse spacing. The pulse sequence introduced in the previous section for the amplification of the oscillations contains by construction an even number of pulses. However, this sequence is symmetric with respect to the middle point $t=\tau$, such that $P(\tau + h) = P(\tau - h)$. Therefore, one could design a pulse sequence containing an odd number of pulses in each of the two halves of the sequence, such that the relative phase is suppressed for each of them. One can express this mathematically, noticing that the integral between $0$ and $T$ for the amplification sequence is just two times the integral from $0$ to $\tau$, $\int_0^T P(t') dt' = 2 \int_0^\tau P(t')dt'$. Now this integral is clearly 0 if one has a number of pulses $N$ between $0$ and $\tau$ that is odd. 

Therefore, we find that with a suitably designed sequence, the pulsed approach of section~\ref{sectionI} can suppress the relative phases introduced by  homogeneous forces acting in the direction of the superposition that are time-independent and independent of the position along the drop. This is due to the fact that for an odd number of pulses both parts of the superposition follow paths whose average displacement in the x-direction across the entire path vanishes. Notice that this strategy is independent of the initial state of the system. With such a pulse sequence the experiment becomes insensitive to misalignments of the setup with the vertical, as gravitational inhomogeneities along the splitting dimension are suppressed. This suggests that, in free-fall experiments, one could tilt the setup on purpose in order to make the particle fall slower, and therefore perform the experiment with a shorter device. This should be beneficial as it would allow the use of shorter magnets, relaxing the requirements for homogeneity of the magnetic field gradient to shorter distances. In the limit where the magnetic field gradient is parallel to gravity, $\beta = \pi/2$, the particle would not fall at all, and the experiment would be equivalent to an experiment performed with a levitated diamond instead of a free-falling one, which would have the significant advantage that a single diamond could then be used repeatedly.

Designing universal pulse sequences that can suppress the effect of higher order potential terms is difficult and may not be possible in general. However, for those higher order contributions that are small and do not distort the oscillation path of the diamond significantly, an extension of the pulse sequence that symmetrises both paths of the interferometer can be designed. Such a pulse sequence should, therefore, be able to suppress the relative phase introduced by potential terms of arbitrary order that are weak and static (or sufficiently slowly varying). The extended sequence is obtained by simply repeating the same pulse sequence with the sign inverted, which is achieved by the application of a $\pi$-pulse in between the two sequences. In this manner the path traveled by spin up of the NV in the second half is that traveled by spin down in the first one, and vice versa, see Appendix~\ref{App:HighOrder}.

\subsubsection{Stochastic homogeneous forces}
Forces that are homogeneous along the oscillation dimension of the diamond but have a magnitude---and/or sign---that changes in time are also to be expected. These can arise for a variety of reasons. In drop experiments, spatial inhomogeneities along the z-axis translate into
time-dependent variations in a frame moving with the center of mass of the nanodiamonds. Due to the moderate velocity of the nanodiamonds these variations will be slow and a moderate number of pulses will suffice to average them out. 

An intrinsic source of time-dependent fluctuations in nanodiamond material is the presence of nuclear and electron spins in the bulk and the surface. Indeed, the flip of one or more of the surface spins, alters the magnetic moment of the diamond and, consequently, the force exerted on it by the magnetic field gradient. Consider an optimistic relaxation time of $\tau = 1000$ s for the surface spins, which indicates that the spins will flip at a rate of $\Gamma = 1/\tau = 1$ mHz,  then for a diamond containing as little as $10000$ surface electron spins, one should expect 10's of flips for each experiment lasting $1$ s. If no counter measures are taken, each spin-flip will lead to a change in the force that is being exerted on the nanodiamond by the magnetic field gradient and hence a stochastic phase that will destroy the interference. 

In order to describe the impact of spin flips and the effect of our pulsed schemes, we adopt a semi-classical approach, where surface spins are always in a projected state of the $\sigma_z$ basis and can randomly fluctuate between its two available states $\{ \ket{\uparrow}, \ket{\downarrow} \}$. Under this approximation one can describe the total instantaneous spin projection in the $z$-direction with a real valued function that fluctuates around its thermal expectation value, $ Z (t) = \expval{\sum^{N_{\rm e}}_i S^z_i}_{\rho^{\rm e}_{\rm th}}+ \eta (t)$, where $N_{\rm e}$ is the number of electronic spins. Here we have introduced the function $\eta(t) = \sum_{i=1}^{N_{\rm e}} X_i(t)$, where $X_i(t)$ represents the state of spin $i$ at time $t$ and is a function that flips stochastically between values $+1$ and $-1$ at a constant rate given by $\Gamma = 1/\tau$, the inverse of the longitudinal relaxation time of the spin. We can model $X_i(t)$ as a random telegraph signal, such that the number of spin flips in the time interval $(t_1, t_2]$ follows a Poisson distribution with parameter $\Gamma (t_2 - t_1)$, and is independent from spin flips in previous intervals. For simplicity we will assume that the spins are initially with equal probability in the states $\ket{\uparrow}$ or $\ket{\downarrow}$. Now, under this assumption and that of no correlation among different spins, we have that $\overline{X_i(t)} = 0$ and $\overline{X_i(t)X_j(t')} = e^{-|t-t'|/\tau} \delta_{ij}$, where the overline indicates averaging over different realisations of the stochastic process. It follows that $\overline{\eta(t) } = 0$ and $\overline{\eta(t)\eta(t')} = N_{\rm e} e^{-|t-t'|/\tau}$, where we have assumed that the flip rate for each spin is the same. 

Therefore, the force on the diamond fluctuates as $ \hbar \gamma_{\rm e} B'  [ \expval{\sum_i S^z_i}_{\rho^{\rm e}_{\rm th}} + \eta (t) ]$, where $\gamma_{\rm e}$ is the electronic gyromagnetic ratio and $B'$ the magnetic field gradient. This modifies our Hamiltonian as
\begin{equation}
\begin{split}
H_x =&\quad \hbar \omega c_t^\dag c_t + \hbar \lambda (c_t + c_t^\dag) P(t) S_z \\
&+ \hbar \tilde \phi P(t)S_z + \hbar \phi \eta (t) P(t)S_z,
\end{split}
\end{equation} 
where $P(t)$ accounts for the pulse sequence on the NV and the ladder operators are now time dependent, $c_t = a + \frac{\tilde F_0 x_0}{\hbar \omega} + \eta (t) \frac{F_0x_0}{\hbar \omega} $. Here, $\phi = \tilde \phi / \expval{\sum_i S_i^z}_{\rho_{\rm th}^{\rm e}} = \Delta_{x_{\rm eq}} F_0/(2 \hbar)$, with $F_0 = \tilde F_0 / \expval{\sum_i S_i^z}_{\rho_{\rm th}^{\rm e}} = \hbar \gamma_{\rm e} B'$. The relative phase introduced by the constant term $\hbar \tilde \phi P(t) S_z$ can be suppressed by a suitable pulse sequence $P(t)$ as described in the previous section. For the stochastically fluctuating term $\hbar \phi \eta (t) P(t) S_z$, however, this will introduce a relative phase between the two states of the NV, which will stochastically vary in time as
\begin{equation}
\label{eq:phase}
\Phi (t) = {\rm mod}_{2 \pi} [2 \phi \int_0^t \eta(s) P(s) ds] - \pi,
\end{equation}
where the function ${\rm mod}_{2\pi} [x]$ gives the remainder with positive sign of $\frac{x}{2 \pi}$, such that $\Phi (t)$ always lies in the interval $[-\pi, \pi]$, and its average is zero. Interestingly enough, the acquired relative phase is independent of the magnetic field gradient. This is because although the force is proportional to the magnetic field gradient the separation distance between the equilibrium positions of the two superposed oscillators is inversely proportional to it, and thus $\phi = \frac{\hbar \gamma_{\rm e}^2 \mu_0}{-\chi_{\rm V} V}$ depends only on the volume of the considered diamond. 

The two superposed states of the diamond pick up a relative phase at a rate $2 \phi \eta(t) P(t)$, however, for a constant function $\eta(t) = \eta_0$ the pulse sequence $P(t)$ makes sure to invert the sign of the rate for adjacent intervals between pulses, such that the phase built up before a $\pi$-pulse is subtracted after it. However, if a spin flip occurs and thus function $\eta(t)$ fluctuates in between two pulses, the accumulated phase is not compensated for in the subsequent interval, and a relative phase is acquired between the two states of the NV. Naturally, this relative phase will be bounded by the interpulse spacing, such that $\phi' \leq 2 \phi \Delta \eta \Delta t$, where $\Delta \eta$ is the variation of function $\eta$ and $\Delta t$ is the interpulse spacing. If additional spin flips occur during the experiment, the phases induced by each of these spin flips will add up. For a $230$ nm radius nanodiamond we find $\phi \sim 10^6$, which indicates that the phase induced by a single spin flip in an experiment lasting half a second will already require a number of pulses on the order of $10^7$ to be suppressed reliably. This sets a daunting perspective on the use of pulse sequences to counteract the fluctuations of surface spins, as one is interested, in order to maximise the separation distance of the superposition, in having a pulse sequence that is resonant with the oscillation frequency of the diamond, which is, typically, several orders of magnitude smaller.

\begin{figure}[htbp]
\begin{center}
\includegraphics[width=\columnwidth]{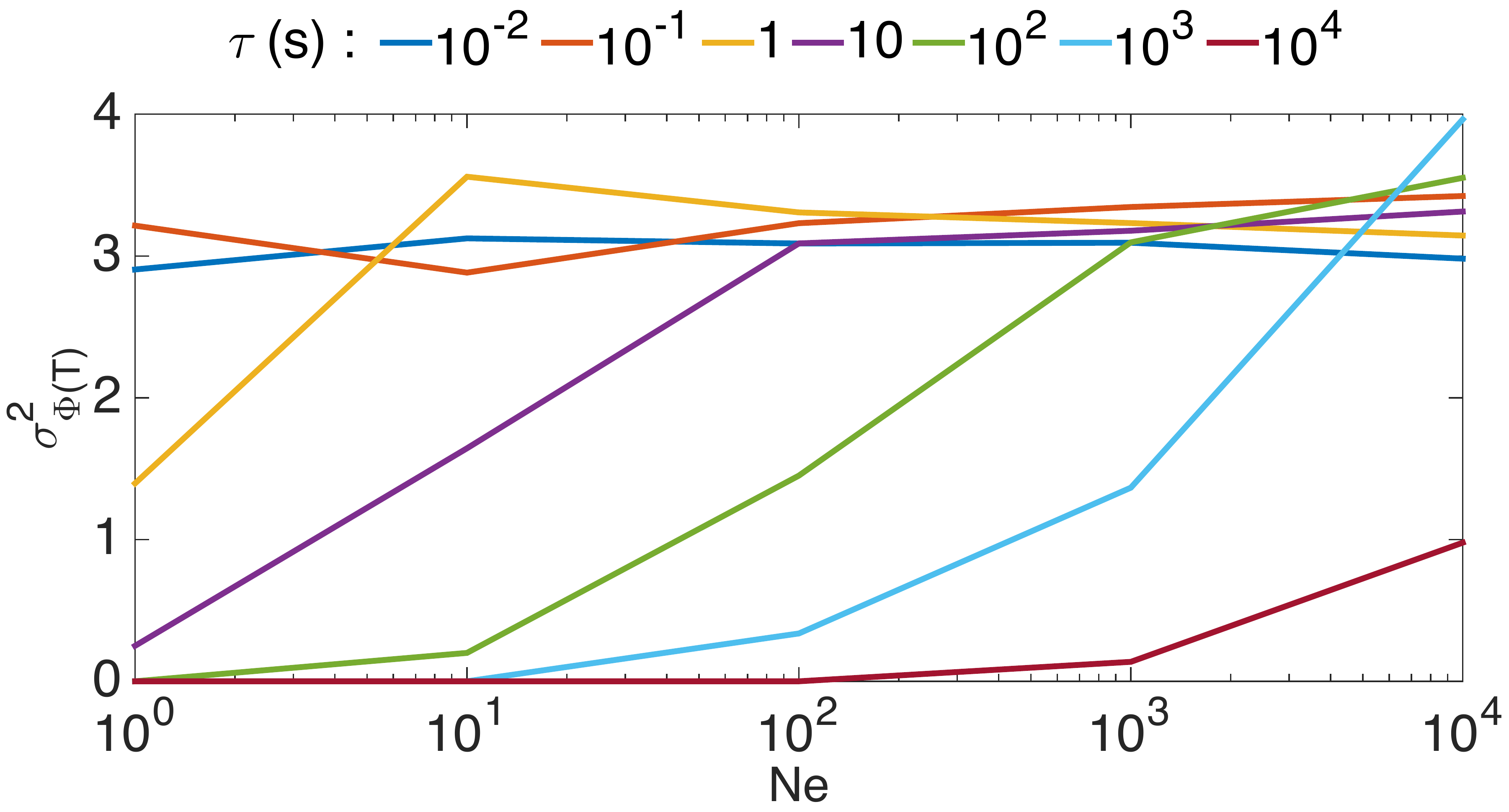}
\caption{{\bf Variance of the relative phase} between the two superposed states of a 230~nm diamond at the end of a protocol of $0.5$~s duration consisting of equally spaced 100 $\pi$-pulses. The x-axis shows the total number of electronic surface spins, and the different colors correspond to different longitudinal relaxation times of these spins. The plot is produced by numerically computing the integral in Eq.~(\ref{eq:phase}) and averaging over 100 realisations of the stochastic function $\eta(t)$}
\label{fig:variance}
\end{center}
\end{figure}

To estimate the impact of such a stochastic force on the visibility of the interference pattern we consider the variance of the relative phase, $\sigma^2_{\Phi (T)} = \expval{\Phi(T) \Phi (T)} - \expval{\Phi (T)}^2$, at the end of an experiment lasting $T = 0.5$~s for a $230$~nm radius diamond. In Fig.~(\ref{fig:variance}) we numerically simulate the stochastic process and compute the variance at the end of a sequence of 100 $\pi$-pulses, averaging over 100 realisations of the stochastic function.  Unsurprisingly, the variance grows with the number of surface spins, as this increases the chances of experiencing spin flips during the protocol. Similarly, surface spins with longer relaxation times show a smaller variance of the phase at the end of the protocol. Notice that for an increasing number of spins, all the setups saturate to a value of the variance that corresponds to that of a homogeneously distributed stochastic variable in the interval $[-\pi,\pi]$, that is $\sigma^2 = \pi^2/3$. All in all, for such a small number of pulses the effect of the sequence in suppressing the build-up of relative phases is negligible, and acceptable variances of the phase are reached only for configurations of the number of spins, $N_{\rm e}$, and relaxation times, $\tau$, that have a low probability of experiencing a spin flip during the experiment.

It is noteworthy to mention that in the presence of a magnetic field gradient, flip-flops between surface spins will also lead to phase fluctuations, as different spatial distributions of the spins are energetically inequivalent. This phase difference will depend on the magnetic field gradient as well as on the spatial orientation of the spin relative to the gradient and, furthermore, the flip-flop rate will decrease with increasing magnetic field gradient. This  renders a detailed analysis of the effect of flip-flops difficult, and hence we relegate it to a future work.

\subsection{Imperfect overlap of the wave functions}

Maximal visibility of the interference requires that the wave packets propagating along the two paths of the matter-wave interferometer are brought back together such that they become indistinguishable in all degrees of freedom, i.e. the positions and momenta of all their constituent particles. This does not, however, require localisation of the translational degrees of freedom to length scales of atomic nuclei but, assuming that the diamond is a rigid body, merely to the minimal size of the wave packet of the center of mass of the diamond in its confining potential. This is given by the extent of the ground state wave packet $x_0 = \sqrt{\frac{\hbar}{2 M \omega}}$. Furthermore, this also requires that rotations around its center of mass are kept in a range that ensures a displacement of the outermost components of the diamond that is smaller than $x_0$. The same requirements should apply to an initial motional state that is thermal, as thermal states are a mixture of coherent states, which have the same width as the ground state, and the overlap requirements should hold for each of the components of such a mixture. In this section we will discuss the physical origin for each of these two sources of imperfect overlap, namely, that in which (1) the centers of mass do not coincide, and that in which (2) the two superposed states of the diamond are rotated with respect to each other. As we will see, the first can occur due to the presence of forces that are inhomogeneous along the two legs of the interferometer, the second due to inhomogeneous torques acting on the diamond. Throughout this section quantities that are given without units are assumed to be in SI units.

\subsubsection{Inhomogeneous forces}

A differential force acting on the superposed states will lead to a relative shift of the centers of mass, such that when reunited these will not perfectly overlap. The relative displacement caused by a force difference $\Delta F$ that acts over a time $t$ on a body of mass $M$ is given by $x_{\rm dis} = \frac{\Delta F t^2}{2M}$. Note that this is only true if the force acts along the dimension for which the motion of the particle is free. For forces acting along the dimensions for which the particle is trapped, such uneven forces will result in a relative displacement of the equilibrium positions by $\Delta x_{\rm eq} = \frac{\Delta F}{M \omega^2}$, which is less restrictive and does not grow in time. To guarantee that the centers of mass are not separated by more than $x_{\rm min}$ the force difference between the two legs of the interferometer   should be bounded by $\Delta F < \frac{2 x_{\rm min} M}{t^2}$. For a 230 nm radius diamond (mass $M=1.8\cdot 10^{-16}$ Kg) and if we assume that the imbalance in the force is present for a duration of $1$ ms---it is reasonable to expect that such a force difference will not be maintained over time as, under the pulse sequence, both superposed states of diamond oscillate across both regions of the setup---and we take $x_{\rm min} = x_0 = 10^{-11}$ m, we find a bound of $\Delta F < 3.6\cdot 10^{-21}$ N. Perturbations in the electric or magnetic fields due to the presence of unwanted electrons or other particles on the surface of the magnets should not be able to generate such a big force difference between the two superposed states. For example, if we take the electric field generated by an electron sitting $100$ microns from the diamond, and we consider that the two paths are separated by $100$~nm, the force difference due to the interaction of the intrinsic electric dipole moment of the diamond with the gradient of the electric field is of the order of $\Delta F \approx 10^{-27}$~N.

A more stringent requirement is put forward by the relative gravitational phase that the two superposed states of diamond can acquire when the relative displacement occurs in the vertical direction. For a difference in altitude $\Delta h$ of the centers of mass, a relative phase is built up that grows in time as $\Delta \phi = \frac{ Mg\Delta h}{\hbar} t$. It follows that for a force difference $\Delta F$ acting during a time t on the vertical axis, phase is accumulated as $\Delta \phi = \frac{g \Delta F}{2 \hbar} t^2 T$, where $T=0.5$~s is the duration of the experiment. Again, if one considers this differential force to act on the system during a millisecond, then $\Delta F \approx 10^{-26}$~N would already introduce a relative phase on the order of rads.

\subsubsection{Rotational dynamics}

In addition to its translational degrees of freedom, a diamond that is either in free fall or levitated can also experience rotation around its center of mass. Rotations of the diamond that are different for each of the two superposed states will lead to a reduction of the visibility of the interference. Additionally, uncontrolled rotations will lead to an unknown Zeeman shift of the NV center, which in turn will result in imperfect pulses. In general, a relative rotation that does not affect adversely the interference must be such that the particles on the surface of the diamond suffer a displacement that is smaller than the spatial extent of their wave function, that is $r \theta \ll x_0$. For a spherical nanodiamond of radius $r = 230$~nm in a trap of frequency $\omega = 1$~kHz we have $x_0 \approx 10^{-11}$~m, which sets a bound on the tolerable relative rotation angles of $\theta \ll 10^{-5}$~rad.
 
The origin of unwanted relative rotations can be manifold. On the one hand, the force separating the two superposed states of diamond acts on the NV center, which, in general, will be located at some distance from the center of mass of the diamond, as schematically shown in Fig.~(\ref{fig:Scheme} c). Evidently, this will lead not only to a translation of the center of mass but to a simultaneous rotation. 

\begin{figure}[htbp]
\begin{center}
\includegraphics[width=\columnwidth]{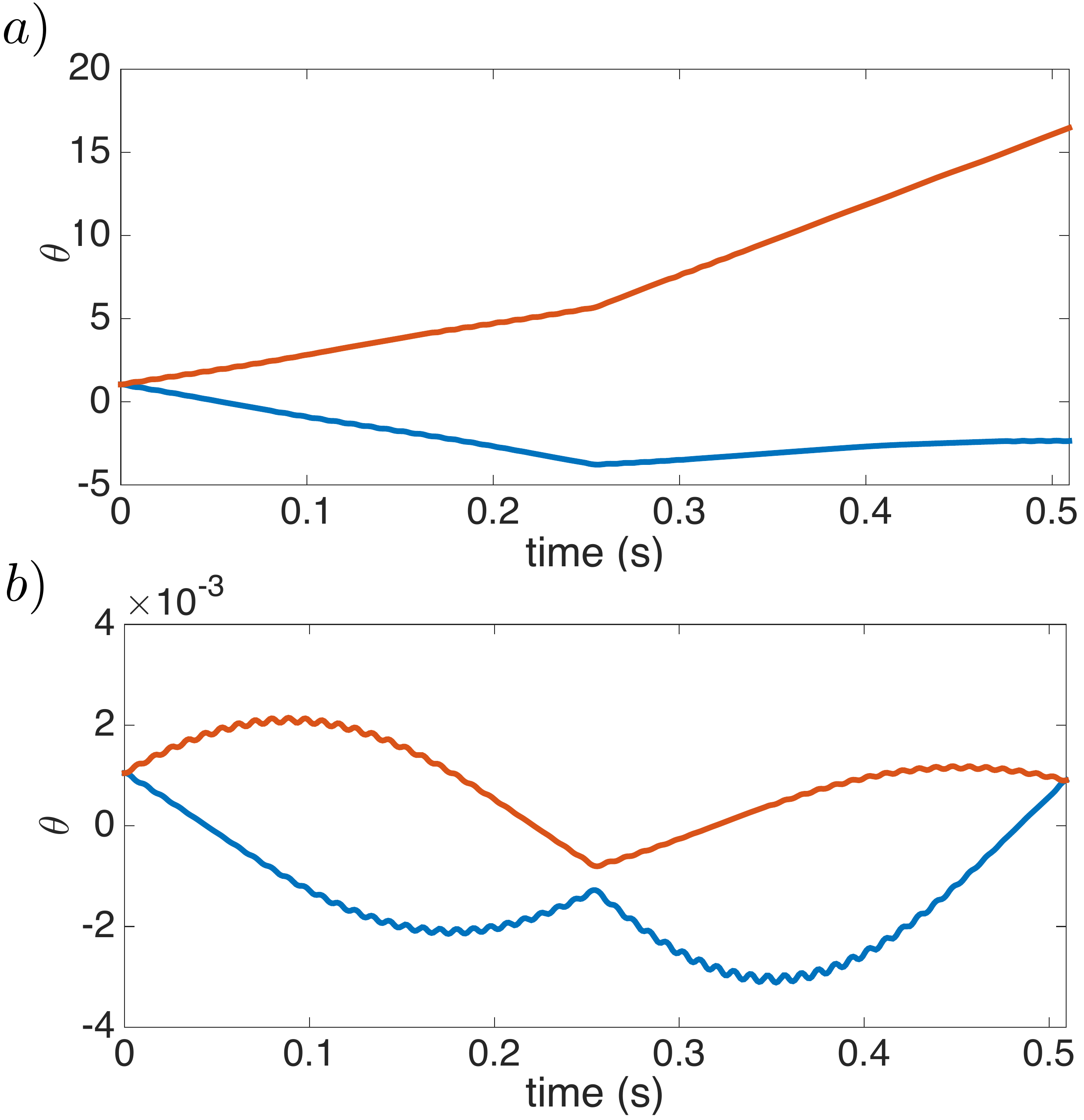}
\caption{ {\bf Rotational dynamics.}  We plot the numerical solution of Eq.~(\ref{eqmotrot}) for a $230$~nm diamond in a $10^4$~T/m magnetic field, which contains an NV center located at a distance from the center of mass that is 80 \% of the radius. Red lines show the dynamics with the NV in state $| + \rangle$ and blue lines with it in state $| - \rangle$. We consider two different initial states: (a) $\theta_0=\pi/3$~rad and (b) $\theta_0 = \pi/3 \times 10^{-3}$~rad. In both cases the initial angular velocity is vanishing. The simulated sequence is the same for both simulations; it lasts approximately 0.5~s and contains in each half 55 pulses, which have a frequency that is twice that of the oscillation of the diamond and therefore is an amplification sequence.}
\label{fig:rotationalDynamics}
\end{center}
\end{figure}

An equation of motion for the rotation angle can be established according to
\begin{equation}
\bm \tau = I \bm{\ddot \theta},
\end{equation}
where $\bm \tau$ is the torque, $I$ the moment of inertia and $\bm{\ddot \theta}$ the angular acceleration of the rotation angle $\theta$. The torque can be expressed in terms of a vector $\bm d_{\text{nv}}$ indicating the position of the NV from the center of mass and the force $\bm F_\text{nv}$ acting on the NV as $\bm \tau = \bm d_{\text{nv}} \times \bm F_\text{nv}$, leading to the following equation of motion
\begin{equation}
\label{eqmotrot}
\ddot \theta  - F(t) q \sin \theta = 0,
\end{equation}
where 
\begin{equation}
q = \frac{3 \hbar \gamma_{\rm e}B'}{2\pi \rho_{\text D} r^4}.
\end{equation}
For the calculation we have assumed that the NV lies at a distance from the center of mass that is $80\%$ of the radius, we have used the moment of inertia of a sphere $I = \frac{2}{5} M r^2$ and the magnetic force acting on the NV spin to be of the form $|\bm F_\text{nv}| = \hbar \gamma_{\rm e} B'$. $F(t)$ was defined in the previous section and is a function oscillating between values $1$ and $-1$ that inverts the direction of the force with each $\pi$-pulse. Equation~(\ref{eqmotrot}) can be numerically solved for a specific pulse sequence and strength of the magnetic field gradient. In Fig.~(\ref{fig:rotationalDynamics}), we show the solution for the case of a $230$~nm diamond in a magnetic field gradient of $10^4$~T/m. We observe that in general, the rotational dynamics of the diamond is not inverted in the second half of the sequence, as is the case with the displacement of the center of mass, leading to relative angles between the two superposed states of diamond at the end of the experiment, which jeopardise the interference.  This is because, unlike the force exerted by the NV on the center of mass, its associated torque is not constant but it depends nonlinearly (sinusoidally) on the rotation angle. In particular, we observe a strong dependence on the initial state. For an initial angle of $\theta = \pi/3$~rad the relative final rotation can be of the order of radians, while for smaller initial angles like $\theta = \pi/3 \times 10^{-3}$~rad this is suppressed under a sequence that contains an odd number of pulses in each of its two halves and null initial angular velocities. This is due to the linearisation of the sine in Eq.~(\ref{eqmotrot}), see Appendix~\ref{appendixRotRef}, and indicates that the same pulse sequence that is used to amplify the oscillation amplitude and cancel relative phases can serve to suppress relative rotations as well, as long as the initial rotation angle is small. In the next section we explain how to trap the rotational degree of freedom in order to keep initial rotations below a desired threshold. Notice that initial rotation angles of $\theta_0 = 0$ will lead to no rotational dynamics, as the torque on the diamond is cancelled. One additional observation is that the quantity $q$ depends on the radius of the particle as $1/r^4$, which suggests that smaller particles should be significantly more susceptible to relative rotations of the diamond. 
The force exerted by the magnetic field on the NV is not the only source of torques. Indeed, as discussed earlier, a neutral nanodiamond containing one negatively charged NV center, has by necessity an intrinsic electric dipole moment of the form $ \bm d = e \bm r_{\text d}$, where $e$ is the charge of the electron and $\bm r_{\text d}$ the vector connecting the NV with the compensating positive charge. In the presence of electric fields, this dipole will also lead to a torque that will rotate the diamond. Variations of the electric field over the two paths of the interferometer lead to relative rotations. In general, the torque exerted by an electric field on a dipole is given $\bm \tau = \bm d \times \bm E$. Therefore the difference in the torque that each of the two states of the diamond would experience is given by $\bm {\Delta \tau} = \bm d \times \bm{\Delta E}$. Assuming that the electric field is uniform over the size of the dipole, we have a relative rotation of $\Delta \theta = \frac{\Delta \tau}{2I} t^2$. Let us consider that the electric field and the dipole form on average an angle of $\pi/4$~rad; this leads to the following estimate
\begin{equation}
    \Delta \theta/t^2 = \frac{3 e r_{\text d} \Delta E}{40 \pi \rho_{\text D} r^5}.
\end{equation}
For example, for a $230$~nm particle, an electric field that differs just by $\Delta E = 10^{-2}$ V/m in the positions of the superposed states of the diamond will already introduce a relative rotation on the order of radians, assuming a protocol of $0.5$ s and $r_{\text d} = 100$~nm. As a reference, the electric field generated by one single electron is given by $E = 14.386 \times 10^{-10} r^{-2}$, where $r$ is the distance from the electron. This shows that the difference in electric field, $\Delta E = \partial_r E|_{r_0} \Delta r$, over a separation distance of $\Delta r = 100$~nm can already be on the order of $0.3$~V/m for a distance of $r_0 = 10 \mu$m from a single electron.

\subsubsection{Trapping the rotational degree of freedom}

Due to the anisotropic polarisability of irregularly shaped diamonds, an externally applied homogeneous electric field will exert a torque on the diamond. As a consequence, the longer dimension of the diamond will try to align with the electric field, leaving its rotational degree of freedom trapped. This mechanism has been experimentally demonstrated~\cite{Li2016b}. 

In the presence of an electric field $\bm E$ a dipole moment will be induced in the diamond of the form
\begin{equation}
\bm d_\text{ind} = \frac{V}{\mu_0 c^2} ( \chi_x E_x \text{\bf e} _x + \chi_y E_y \text{\bf e}_y + \chi_z E_z \text{\bf e} _z  ),
\end{equation}
where $\{ \text{\bf e}_i \}$ are unit vectors pointing in the directions of Cartesian coordinates in the reference frame of the diamond, and $\chi_i$ is the electric susceptibility in direction $i$. Let us consider an electric field vector laying on the x-y plane and forming an angle $\theta$ with the x-axis of the diamond, such that $\bm E = E \sin \theta \text{\bf e} _x + E \cos \theta \text{\bf e}_y $. The torque exerted by the electric field on the diamond will be 
\begin{equation}
\begin{split}
\bm \tau_\text{ind} &= \bm d_\text{ind} \times \bm E  \\
&= \frac{V E^2}{ 2 \mu_0 c^2} (\chi_x - \chi_y) \sin \theta \cos \theta \text{\bf e}_z.
\end{split}
\end{equation}
The difference in electric susceptibility for the different coordinates of diamond will depend on the specific shape of the diamond. For an estimation of the achievable torque we will assume $\chi_x - \chi_y = 0.3$ \cite{Li2016b}. For small angles one can take $\sin \theta \cos \theta \approx \theta$, such that for a particle with the volume that equals that of a sphere of radius $230$ nm the torque induced by the electric field is of the form
\begin{equation}
\label{indTor}
\tau_\text{ind} \approx \frac{V}{2 \mu_0 c^2} 0.3 E^2 \theta \approx 6.76\times 10^{-32}\ E^2 \theta.
\end{equation}
As already discussed, an additional torque will be exerted by the same field on the intrinsic dipole moment of the diamond---the one formed by the negative charge of the NV and its compensating charge---, which assuming that the charges sit $100$~nm apart from each other has a magnitude
\begin{equation}
\label{intTor}
\tau_\text{int} = e d_\text{nv} E \approx 10^{-26} E,
\end{equation}
in the most unfavourable case, when the dipole is perpendicular to the applied field. This tells us that the diamond will be locked in the angle for which $\tau_\text{ind} + \tau_\text{int} = 0$. Small field inhomogeneities on top of the applied field will exert a torque that will try to pull the diamond out of this equilibrium position. For example, an electric field of uncontrolled origin with a magnitude of $E_\text{unc} = 10$~V/m, will generate a torque on the internal electric dipole of the NV of the order $\tau_\text{int} = 10^{-24}$ J. An applied electric field of $E_\text{ind} = 10^6$~V/m would compensate with an induced torque $\tau_{\rm ind} \approx 10^{-20} \theta$, which would keep rotations below $\theta < 10^{-4}$~rad. Torques due to the magnetic field gradient acting on an NV that sits off the center of mass of the diamond at a distance $80 \%$ of its radius, which we assume to be $230$~nm, are on the order of $\tau = 10^{-30} B'$. If we consider a working magnetic field gradient of $10^4$~T/m, rotations of such an origin can be restricted to the same degree with a smaller electric field on the order of $10^5$~V/m.

Regarding thermal excitations of the rotational degree of freedom, one can associate a potential energy to the torque induced by the applied electric field of the form
\begin{equation}
U_\text{rot} = \frac{V E^2}{2 \mu_0 c^2} (\chi_x - \chi_y) \sin^2 \theta,
\end{equation}
such that $\bm \tau_\text{ind} = \partial_\theta U \text{ \bf e}_z$. Assuming small angles, we can estimate the maximum reachable rotation for a given energy $U$ as
\begin{equation}
\theta_\text{max} = \sqrt{\frac{2 \mu_0 U}{V(\chi_x - \chi_y)}} \frac{c}{E}.
\end{equation}
Therefore, at temperature $T$, a particle of radius $230$ nm should experience thermal rotations not bigger than
\begin{equation}
\begin{split}
\theta_\text{max} &= \sqrt{\frac{2 \mu_ 0 k_\text{B}T}{V (\chi_x - \chi_y)}} \frac{c}{E} \\
&= 1.4 \times 10^4 \frac{\sqrt{T}}{E}.
\end{split}
\end{equation}
This indicates that an applied electric field of $10^6$ V/m would keep the rotation angles of a particle whose rotational degrees of freedom are cooled to a temperature of 1 K below $10^{-2}$~rad. Note that thermal rotations do not represent relative angles between the two superposed states of diamond, but the range of angles that the initial state can take. As discussed in previous sections, keeping the initial rotation angles small, allows us to linearise the equations of motion of the rotational degrees of freedom, which leads to the pulse sequence refocusing not only the position but also the rotation of the nanodiamond.

\subsubsection{The NV in the presence of an electric field}
If an electric field is applied in order to trap the rotational degrees of freedom, one should take into account that the Hamiltonian of the NV will be transformed into
\begin{equation}
\label{NVelect}
\begin{split}
H_\text{NV} ={} &\hbar (D + d^{||}E_z) S_z^2 + \hbar \gamma_{\rm e} \bm S \bm B\\
&- \hbar d^{\bot} [E_x (S_x S_y + S_y S_x) + E_y (S_x^2 - S_y^2)]\\
= {}&\hbar (D + d^{||}E_z) S_z^2 + \hbar \gamma_{\rm e} \bm S \bm B \\
&+ \hbar (\Omega_\text{E} \ketbra{+}{-} + \Omega_\text{E}^* \ketbra{-}{+}),
\end{split}
\end{equation}
where $d^{\parallel}$ and $d^{\bot}$ are the axial and non-axial components of the permanent ground-state dipole moment, which have measured values of $d^\parallel = 0.17$~Hz~m/V and $d^\bot = 3.5 \times 10^{-3}$~Hz~m/V, and we have defined an electric field induced Rabi frequency $\Omega_\text{E} = d^\bot (E_y + i E_x)/2$. The presence of continuous rotations between states $\ket{+}$ and $\ket{-}$ is undesired as it will lead to an attenuation of the reachable separation distances between the two superposed states of the diamond. Intuitively, this is because in the interaction picture with respect to such a term, operator $S_z$ in Hamiltonian~(\ref{HamX}) will acquire a time dependence of the form $\lambda S_z (a + a^\dag) \rightarrow \lambda (\ket{+}_x\bra{-}_xe^{-i\Omega_\text{E}t} + \text{H.c.})(a + a^\dag)$, where $\ket{\pm}_x$ are eigenstates of operator $\sigma_x$---here, $\sigma_x$ is a Pauli operator defined in the subspace of the spin-1 system formed by states $\ket{\pm}$. This time dependence will average the effect of the displacement and effectively reduce it.

In order to avoid this, instead of employing a static electric field to trap the rotational degrees of freedom, one could use a field oscillating at frequency $\alpha$ such that $\Omega_\text{E} \rightarrow \Omega_\text{E} \cos \alpha t$. Now if $\alpha \gg \Omega_\text{E}$ the last term in Eq.~(\ref{NVelect}), can be neglected under a rotating wave approximation. Roughly, for an electric field of $10^5$~V/m we find $\Omega_\text{E} = 3.5 \times 10^2$~Hz, so that if the electric field oscillates at $10$~MHz, the conflicting term would introduce an effective Stark shift with magnitude $\Omega_\text{E}^2/\alpha \approx 10^{-3}$~Hz, which over the time of a second should introduce a phase smaller than $10^{-3}$~rad. For higher oscillation frequencies this would only improve, for example if the field from a linearly polarised laser is employed, which oscillates at THz frequencies, we have that $\Omega_\text{E}^2/\alpha \approx 10^{-8}$, and its effect should be negligible.

\section*{Conclusion}

The use of nanodiamonds for matter-wave interferometry is a technically demanding task. Here, we have offered an analysis of experimentally relevant aspects to serve as a roadmap in the pursuit of such a goal. On the basis of our findings, we have modified previously proposed experimental protocols by incorporating a pulse sequence that effectively realises dynamical decoupling of motional degrees of freedom  and can reduce the sensitivity of the setup to a broad range of experimental imperfections, therefore relaxing the technical requirements. With such a protocol we have made the experiment insensitive to the presence of diamagnetic forces, misalignments of the setup, static electric fields or the net magnetic moment of the diamond. These protocols may also find application in more general matter-wave interferometry schemes. Our pulse sequence serves also as a dynamical decoupling sequence that extends the coherence time of the NV center. We have explored the impact of rotational dynamics and found that, for initial angles that are small, our sequence can also refocus relative rotations between the two superposed states of diamond. In order to keep rotations below a certain threshold, we have proposed to lock the rotational degree of freedom with homogeneous oscillating electric fields. On the other hand, we have shown that the presence of Casimir-Polder forces imposes a limit on the distance between the diamond and the magnets and, consequently, on the employable magnetic field gradients. This will have an effect on the oscillation frequency of the diamond and on the number of pulses that one can apply, since these have to be resonant with the oscillations of the diamond. The most challenging aspect is the impact that fluctuating electronic spins from the dangling bonds at the surface of the diamond can have on diamond matter-wave interferometry. Experimental efforts in surface passivation and moving to shorter experimental times seem essential. All in all, our work brings the possibility of matter-wave interferometry with nanodiamonds closer to reality by alleviating some of the experimental requirements and pointing to the needed technical improvements in order to reach it.  

\section*{Aknowledgments}

M. B. P. and J. S. P. acknowledge support by the ERC Synergy grant BioQ (Grant No. 319130), the EU projects HYPERDIAMOND and AsteriQs, the QuantERA project NanoSpin, the BMBF project DiaPol, the state of Baden-W\"urttemberg through bwHPC, the German Research Foundation (DFG) through Grant No. INST 40/467-1 FUGG, and the Alexander von Humboldt Foundation through a postdoctoral fellowship. G. W. M. is supported by the Royal Society.

\appendix

\section{Diamagnetic forces on extensive objects}
A correct treatment of diamagnetic forces should consider variations of the magnetic field over the body of the diamond, instead of treating it as a point particle. Therefore, one should consider a force of the form 
\begin{equation}
\bm F = \frac{\chi_\text{V} B'^2}{\mu_0}\int_\text{V} d\text v \ x \text{ \bf e}_x,
\end{equation} 
where we are considering a magnetic field gradient along the $x$-axis only, and the integral is over the volume of the diamond. Let us assume that the particle is sitting in its equilibrium position such that $\int_\text{V} d\text v \ x = 0$. Notice that in general the center of mass will not necessarily sit on the zero of the magnetic field. For an arbitrarily shaped diamond one can write an integral
\begin{equation}
\int_\text{V} d\text v\  x = \int_{-\infty}^{\infty} dx\ x f(x), 
\end{equation}
where $f(x)$ is a function that gives the area of the section of the body with the $zy$-plane in position x, such that $\int_{-\infty}^\infty dx f(x) = \text V$, where V is the volume of the diamond. Now, if the body is displaced a distance $\Delta x$ in the $x$ direction, we have that the body will feel a force
\begin{equation}
\bm F = \frac{\chi_\text{V} B'^2}{\mu_0}\ \int_{-\infty}^{\infty} dx\ x f(x-\Delta x) \text{ \bf e}_x.
\end{equation}
Under the change of variable $y=x-\Delta x$ we get that 
\begin{equation}
\begin{split}
\bm F &= \frac{\chi_\text{V} B'^2}{\mu_0}\ \int_{-\infty}^{\infty} dy\ (y+\Delta x) f(y) \text{ \bf e}_x\\
&= \frac{\chi_\text{V} B'^2 V}{\mu_0}\ \Delta x,
\end{split}
\end{equation}
where we have used $\int_{-\infty}^{\infty} dy\ y f(y) = 0$ and $\int_{-\infty}^{\infty} dy\ f(y) = \text V$. Therefore, we have that the diamagnetic force is the same as if we treat the particle as a point particle, with the only distinction that the equilibrium position of the diamond will not necessarily correspond to that in which the magnetic field is 0 at the center of mass.

\section{Potential terms of order higher than linear}
\label{App:HighOrder}

We consider a potential term of uncontrolled origin $V(x,y,z)$, which is time independent and has an arbitrary spatial form. Notice that in free-fall experiments time independence translates into homogeneity along the falling dimension. We assume such a term to be small compared to the parameters governing the harmonic motion of the diamond, $ V(x,y,z)/\hbar \ll \omega, \lambda$. In this case it has a negligible impact on the dynamics of the center of mass of the diamond, but it can still introduce a detrimental relative phase between the two paths of the interferometer, which travel different regions of the potential. Under these considerations, the phase built up in each of the paths of the interferometer can be expressed as
\begin{equation}
    \phi_\pm (t) = 1/\hbar \int_0^t dt' \expval{U_\pm^\dag (t',0) V U_\pm (t',0)},  
\end{equation}
where labels $\pm$ indicate the path and $U_\pm(t,0)$ are the corresponding unitary evolution operators from time $0$ to $t$ in the perturbation free case. The time dependence is expressed in the Heisenberg picture, and thus the expectation value is taken over the initial state of the system. Even if $U_\pm (T,0) = \mathbb{I}$ and both paths come to their initial position at the end of the sequence, in general, we will find that the built up phases are inequivalent, $\phi_+ (T) \neq \phi_-(T)$. This is because the two paths are different, and therefore the integrals of the potential function over these two paths take distinct values, see Fig.~(\ref{fig:Scheme}b) in the main text. In order to heal this we suggest to double the protocol with a sign inversion of the pulse sequence in the second half---that is, introducing an additional $\pi$-pulse at time $t=T$ and repeating the whole sequence---, such that the state following path + in the first half follows path - in the second half, and vice versa. In this manner we get that at time $t=2T$ both paths are now equivalent and therefore $\phi_+ = \phi_-$, cancelling relative phases coming from weak static potentials. Formally, we have that $U_\pm (T + t,0) = U_\mp(t,0)$ with $t \leq T$, given that $U_\pm(T,0)=\mathbb{I}$, and thus
\begin{align}
\begin{split}
\phi_\pm (2T) &= 1/\hbar \int_0^{2T} dt' \expval{U_\pm^\dag (t',0) V U_\pm (t',0)} \\
      &= 1/\hbar \int_0^T dt'  \left( \expval{U_\pm^\dag (t',0) V U_\pm (t',0)} \right. \\ 
     &\hspace{75pt} \left. + \expval{U_\mp^\dag (t',0) V U_\mp (t',0)} \right) . \end{split}
\end{align}
And therefore, such an extended sequence suppresses relative phases introduced by weak, static potential terms of arbitrary order. 

\section{Inhomogeneities of the magnetic field gradient}
We consider here a perturbation to the magnetic field gradient that is quadratic in the position, that is $B(x) = B'x + B''x^2$. This will have two effects, on the one hand, the perturbation should introduce modifications to the diamagnetic force and, on the other hand, to the force on the NV. In particular, we find perturvative Hamiltonian terms of the form
\begin{equation}
    V_{\rm per} = - \frac{\chi_{\rm V} {\rm V}}{2 \mu_0}(B'B''x^3 + B''^2x^4) - \hbar \gamma_{\rm e} B'' x^2 S_z.
\end{equation}
The presence of such terms is, naturally, detrimental to the interferometer as they will entail the appearance of relative phases between the two different paths as well as a distortion of the paths that could result in their imperfect overlap at the end of the protocol. In the case in which the effect of these terms is weak as to not distort the paths the relative phase they introduce can be suppressed with an extended pulse sequence as that described in Appendix~\ref{App:HighOrder}. For that, a general requirement should be that $B''\expval{x} \ll B'$ at all times. 

\section{Impact of Casimir-Polder Potential}
\label{app:Casimir}
As the smallest acceptable separation of the nanodiamonds from the magnetic tip limits the achievable magnetic field gradient, one needs to consider interactions between a neutral body and a magnetic surface. The strength of the interaction depends on the nature of the magnetic material. Here we assume it to be a purely dielectric material, which results in the weakest interaction with the dielectric nanodiamond, the Casimir-Polder force. The Casimir-Polder potential between two dielectric spheres of radius $R_1$ and $R_2$ separated by a distance $r$ is given by 
\begin{equation} \label{CasimirPolder}
    V(r) = \frac{23\hbar c R_1^3 R_2^3}{4\pi r^7}\left(\frac{\epsilon_r-1}{\epsilon_r+2}\right)^2,
\end{equation}
and thus its gradient is
\begin{equation} \label{CasimirPolderGradient}
    |\frac{dV}{dr}(r)| = \frac{161\hbar c R_1^3 R_2^3}{4\pi r^8}\left(\frac{\epsilon_r-1}{\epsilon_r+2}\right)^2.
\end{equation}
The largest tolerable gradient in the Casimir-Polder potential is directly related to the smallest uncertainty in the position of the nanodiamond that can be achieved. Hence our goal is to maximise 
the magnetic field gradient for a given gradient of the Casimir-Polder potential for a spherical magnetic sphere of radius $r_0$. For given $|\frac{dV}{dr}(r)|$ we have
\begin{equation}\label{eq34}
    r = \left[C|\frac{dV}{dr}(r)|^{-1} r_0^3\right]^{1/8},
\end{equation}
with $C = \frac{161\hbar c R_1^3}{4\pi}\left(\frac{\epsilon_r-1}{\epsilon_r+2}\right)^2$ and the magnetic field gradient is 
\begin{eqnarray}
    \frac{dB}{dr} &=& \frac{\mu_0 M}{r_0}\frac{1}{(1 + r/r_0)^4}\\
    &=& \frac{\mu_0 M}{(r_0^{1/4} + r\cdot r_0^{-3/4})^4}.
\end{eqnarray}
Substituting Eq. (\ref{eq34}) to eliminate $r$ and then finding the $r_0$ that maximises the gradient by minimising the denominator as a function of 
$r_0$ yields
\begin{equation}
    r = r_0 = \left(\frac{3}{2}\right)^{8/5}\left[C|\frac{dV}{dr}(r)|^{-1} \right]^{1/5}.
\end{equation}
For a position uncertainty $\Delta r$ and a drop time of $t=0.5$ s, the accumulated phase uncertainty 
in a given gradient of the Casimir-Polder potential is 
\begin{equation}
\Delta \phi = \frac{|\frac{dV}{dr}(r)|\Delta r}{2\hbar}.
\end{equation}
In order to preserve the possibility for interference fringes we require $ 1\gg \Delta \phi$ from which it follows
\begin{eqnarray}
    \frac{dB}{dr} &=& \frac{\mu_0 M}{16 \left(\frac{3}{2}\right)^{8/5} \left[C|\frac{dV}{dr}(r)|^{-1}\right]^{1/5}} \\
    &\ll& \frac{\mu_0 M}{16 \left(\frac{3}{2}\right)^{8/5} C^{1/5}} \left(\frac{2\hbar}{\Delta r}\right)^{1/5}.
\end{eqnarray}
Hence, if we assume a small magnetic tip with a trapped flux of $\mu_0M=1$ T, a radius of the nanoparticle of $R_1=1\mu$m, $\epsilon_r=5.7$ we have $C= 1.5\cdot 10^{-43}$. If we assume 
that we can control the position of the nanoparticle to $\Delta r = 10^{-10}$ m we find
\begin{equation}
    \frac{dB}{dr} \ll 220 \text{ T/m.}
\end{equation}
For $R_1=100$ nm we find
\begin{equation}
    \frac{dB}{dr} \ll 877 \text{ T/m.}
\end{equation}
These estimates are generous, as typically the magnet will not merely be a sphere but will have a
holder or may be spatially extended along the direction of the drop. This may reduce the right hand 
side by a moderate factor. 

\section{Rotational refocusing}
\label{appendixRotRef}
As discussed in the main text, the force exerted by the magnetic field on the NV will result in a torque on the diamond, as long as the NV is not located exactly at the center of mass. We define the rotation angle as the angle between the direction of the gradient---or equivalently, the direction of the force acting on the NV---and the position vector of the NV as measured from the center of mass of the diamond, see Fig.~(\ref{fig:Scheme}c) in the main text. When the NV is put in a superposition, each of the two superposed states will feel opposite torques and a relative angle will grow between the two. As the equations of motion of the rotational dynamics are nonlinear in the rotation angle, this will lead to a net relative angle at the end of the protocol, which will destroy the interference pattern if no countermeasures are taken. In this appendix we will show that for small initial rotations and angular velocities, a suitable pulse sequence will cancel the relative rotations between the two superposed states of the diamond, such that at the end of the protocol, when the centers of mass are reunited the orientations of the two superposed states of diamond will match. The angle of the final state, however, need not be that of the initial state.

If the rotation angle is small throughout the whole protocol, the equation of motion in Eq.~(\ref{eqmotrot}) can be linearised in $\theta$ such that it looks like
\begin{equation}
    \ddot \theta - F(t) q \theta = 0.
\end{equation}
One can rewrite such a second order differential equation in the form of two coupled linear differential equation as
\begin{equation}
\label{eq:rotlinear}
 \frac{d}{dt} \left[\begin{array}{c} \theta \\ \dot \theta \end{array} \right] = \left( \begin{array}{cc} 0 & 1 \\ F(t)q & 0 \end{array} \right) \left[\begin{array}{c} \theta \\ \dot \theta \end{array} \right].
\end{equation}
Since $F(t)$ is a stepwise function switching between values $1$ and $-1$, the time-dependent differential equation~(\ref{eq:rotlinear}) can be broken down into a sequence of time-independent differential equations of the form
\begin{equation}
\label{eq:seqrot}
 \frac{d}{dt} \left[\begin{array}{c} \theta \\ \dot \theta \end{array} \right] = A_\pm \left[\begin{array}{c} \theta \\ \dot \theta \end{array} \right],
\end{equation}
where
\begin{equation}
    A_\pm = \left( \begin{array}{cc} 0 & 1 \\ \pm q & 0 \end{array} \right),
\end{equation}
such that the rotational dynamics alternates between equations of motion ruled by $A_+$ and $A_-$ with every pulse of the sequence. Equation~(\ref{eq:seqrot}) has an analytic solution of the form
\begin{equation}
    \left[\begin{array}{c} \theta \\ \dot \theta \end{array} \right]_t = e^{A_\pm (t-t_0)} \left[\begin{array}{c} \theta \\ \dot \theta \end{array} \right]_{t_0}.
\end{equation}

We are concerned with pulse sequences that contain an odd number of pulses in the first half of the sequence---that is an even number of intervals---, an avoided pulse in the middle, and the inverse sequence of pulses in the second half, such that the sequence contains an even number of pulses in total with the same interpulse spacing except for the central interval which should be double of the rest, see Fig.~(\ref{fig:Scheme}a). Such a sequence cancels relative phases introduced by linear potentials, and will also cancel relative rotations, as we will show now. For a generic pulse sequence with the just described structure, the time evolution of the rotation angle corresponding to state up of the NV is described by operators $X^+_1$ and $X^+_2$ acting on the initial state as $X^+_2 X^+_1 \left[ \theta; \dot \theta \right]_{t_0}$, where
\begin{align}
\label{eq:pulseseq}
    X^+_1 &= e^{A_- \Delta t} e^{A_+ \Delta t} ... e^{A_- \Delta t} e^{A_+ \Delta t}\\
    X^+_2 &=  e^{A_+ \Delta t} e^{A_- \Delta t} ... e^{A_+ \Delta t} e^{A_- \Delta t}
\end{align}
represent the first and second halves of the sequence, respectively, and $\Delta t$ is the interpulse spacing. On the other hand, the rotational dynamics of the state associated with spin down of the NV follows the opposite dynamics, where every interval associated to $A_+$ for the state up follows the dynamics of $A_-$ for state down and vice versa, that is
\begin{equation}
\label{eq:pulseseqrot}
    X_1^- = X_2^+ \qquad  X_2^- = X_1^+.
\end{equation}
Notice that Eq.~(\ref{eq:pulseseqrot}) is only true when each half of the sequence contains an odd number of pulses. We want to show that for initial states with null angular velocity, $\left[ \theta; 0 \right]_{t_0}$, both evolutions lead to the same rotation angle at the end of the sequence. Notice that for an initial state of the form $\left[ \theta; 0 \right]_{t_0}$, the final angle will be determined solely by the diagonal of the evolution operator, and therefore we only need to show that the diagonals of $X_2^+X_1^+$ and  $X_2^-X_1^- = X_1^+X_2^+$ are the same. In general, the diagonal of the product of two matrices $A$ and $B$ is different from the diagonal of their commuted product, diag($AB$)$\neq$diag($BA$). However, in the particular case when $A$ and $B$ are $2 \times 2$ matrices with equal anti-diagonal elements, we find that diag($AB$) = diag($BA$). Thus, if we show that the anti-diagonals of $X_1^+$ and $X_2^+$ are the same we will have proven our point. Indeed, this is the case: following the construction of matrices $X_1^+$ and $X_2^+$ in Eq.~(\ref{eq:pulseseq}) one can see that 
\begin{equation}
\label{eq:antidiag}
    X_2^+ = \sigma_x (X_1^+)^{\rm T} \sigma_x,
\end{equation}
where $\sigma_X = [0 1; 1 0]$ and $^{\rm T}$ indicates the transpose. In the form of Eq.~(\ref{eq:antidiag}) it becomes evident that $X_1^+$ and $X_2^+$ have the same anti-diagonal elements, as the effect of the $\sigma_x$ matrices cancels that of the transposition, leaving the anti-diagonal elements unchanged. Thus, both sequences result in an evolution operator with the same diagonal elements, and therefore the final rotation angle for each of the spin states is the same when the initial angular velocity vanishes and the sequence contains an odd number of pulses in each of its two halves.

\section{Tests of quantumness of gravity}
\begin{figure}[htbp]
\begin{center}
\includegraphics[width=\columnwidth]{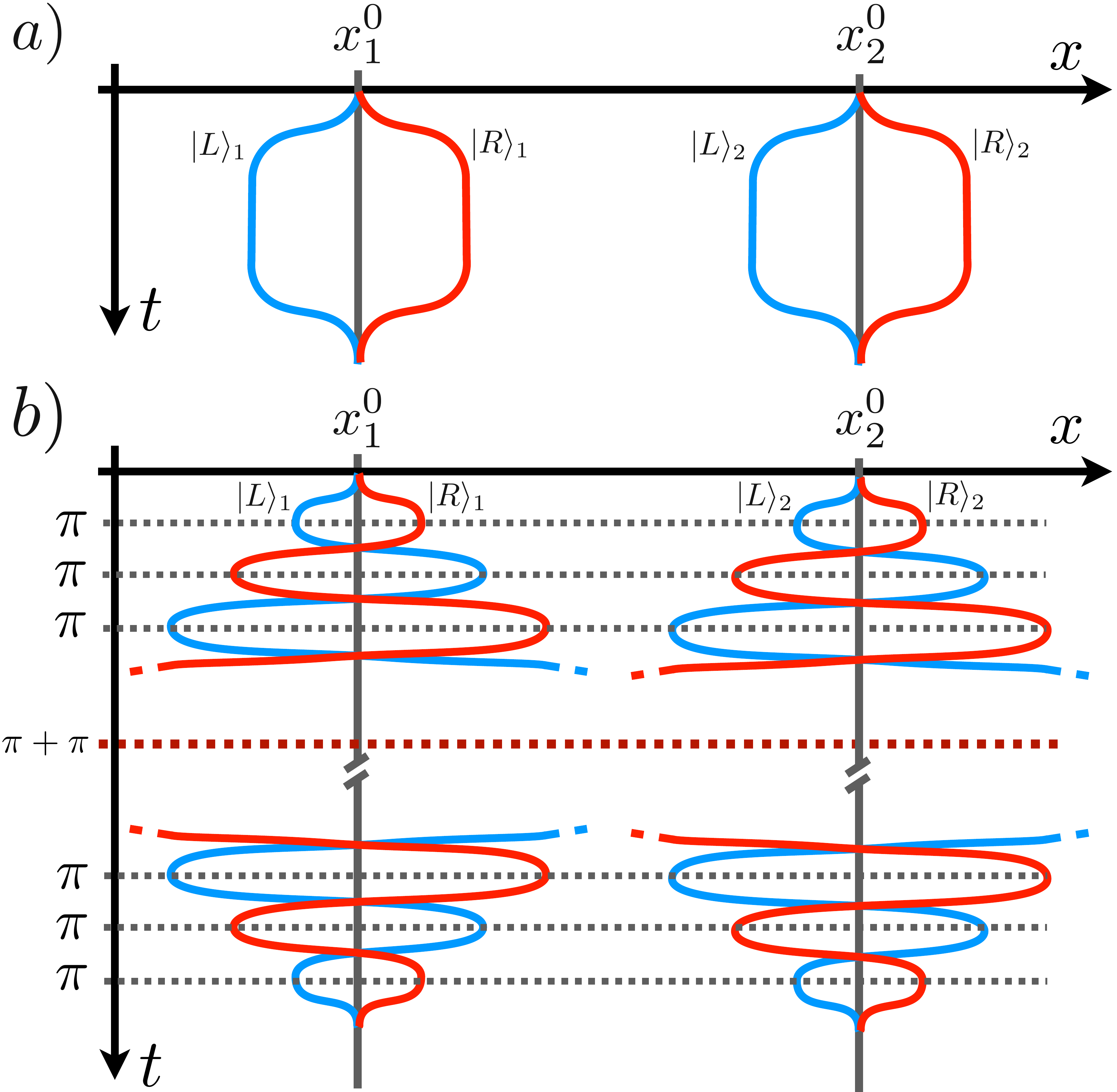}
\caption{ {\bf Interferometer for testing quantumness of gravity.}  (a) shows the original setup presented in~\cite{Milburn2017} where no motional dynamical decoupling is applied. (b) shows the same setup incorporating motional dynamical decoupling.}
\label{FigDoubleInterferometer}
\end{center}
\end{figure}

In a recent work a set-up as the one sketched in Fig.~(\ref{FigDoubleInterferometer}a) has been suggested as a means to test the quantum character of the gravitational interaction \cite{Milburn2017} by measuring gravitationally mediated entanglement between 
the two interferometers. This entanglement arises as the initial state of the interferometer $|LL\rangle+|LR\rangle+|RL\rangle+|RR\rangle$
evolves into $e^{i\phi_{LL}}|LL\rangle + e^{i\phi_{LR}}|LR\rangle + e^{i\phi_{RL}}|RL\rangle + e^{i\phi_{RR}}|RR\rangle$ in the course of the interaction. Taking the parameters of \cite{Milburn2017} one finds distances $d_{LL}=d_{RR}=250\ \mu$m, $d_{LR}=700\ \mu$m and $d_{RL}=200\ \mu$m and, up to a global phase, the wave function $|LL\rangle + e^{i(\phi_{LR}-\phi_{LL})}|LR\rangle + e^{i(\phi_{RL}-\phi_{LL})}|RL\rangle 
+ |RR\rangle$, where $\phi_{LR}-\phi_{LL} \ll \phi_{RL}-\phi_{LL}$ for the gravitational interaction so that the state is
entangled. The relative phase $\Delta\phi_{grav} = \phi_{RL}-\phi_{LL}$ scales as $Gm_1m_2 t_{exp}/(\hbar d_{RL}) \sim 1.6$. It is important 
to note, however, that other forces may also induce entanglement. Notably, in \cite{Milburn2017} a magnetic field gradient of $10^6$ T/m is assumed and the $1\ \mu$m radius of the diamonds implies that they will also experience a significant magnetic field and hence an 
induced dipole moment which scales as $\mu_{ind}=V\chi_V B/2\mu_0$. As a consequence the magnetic dipolar interactions between the 
induced dipole moments in the different arms of the interferometer will also lead to entanglement. Indeed, the phase $\Delta\phi_{dip} 
= \phi_{RL}-\phi_{LL}$ will scale as $\hbar^{-1}\mu_0/(4\pi d_{RL}^3)\mu_{ind}^2 t_{exp}$ which for the chosen parameters will be $\sim 3\cdot 
10^3$ for $B=1$ T. Hence, a much larger distance $d_{RL} \sim 0.1$ m is required to make the contribution of the induced magnetic dipole moment 
small compared to that due to gravity. For such distances, all other things unchanged, the relative phase will be of order $10^{-3}$ and 
the resulting entanglement very small indeed. Furthermore, the large spatial separations that would have to be envisaged in turn imply 
a very significant sensitivity to perturbing effects that have been described in the main text. Beyond those effects, notably, gravitational
interactions with distant masses would have a very important effect. Indeed, a mass of $M=1$ mg at a distance of $10^3$ m would already
impose a relative phase of order $10^{-3}$, which, if uncontrolled, would already be capable of erasing the entanglement thus making the 
effect unobservable. It seems unlikely that the environment of such an apparatus could be controlled to a degree sufficient to eliminate
these deleterious effects. Hence, again, motional dynamical decoupling will be essential adding phase stability to the interferometer. 
Here it is necessary that motional dynamical decoupling is applied to both arms of the interferometer in a correlated fashion so
as to average out decoherence and imperfections from the environment while preserving the phase accumulation due to direct interaction
between the interferometers (see Fig.~( \ref{FigDoubleInterferometer}b) for the envisaged type of paths). Indeed, a direct computation
shows that the direct interaction between the two interferometers, e.g. due to gravity or induced dipoles, is only slightly reduced
while slow external perturbations, notably those linear in position, are averaged out. Nevertheless, despite such improvements the
realisation of tests of the quantum character of the gravitational force is very challenging indeed and appears to require, if possible
at all, a concerted and longlasting effort not dissimilar to that leading to the creation of LIGO, in which material design, quantum control and signal processing need to be driven jointly to new levels of performance.

\end{document}